\documentclass[12pt]{article}

\usepackage{arxiv}
\usepackage{diagbox}
\usepackage{multirow}
\usepackage[utf8]{inputenc} 
\usepackage[T1]{fontenc}    
\usepackage{hyperref}       
\usepackage{url}            
\usepackage{booktabs}       
\usepackage{amsfonts}       
\usepackage{nicefrac}       
\usepackage{microtype}      
\usepackage{lipsum}
\usepackage{graphicx}
\usepackage{amsmath}
\usepackage{amsthm}
\usepackage[all]{xy}
\usepackage{array}
\usepackage[toc,page]{appendix}
\usepackage{tabularx}
\usepackage[affil-it]{authblk}
\usepackage{algorithm}
\usepackage{algpseudocode}
\usepackage{dsfont}

\newtheorem{prop}{Hypothesis}

\theoremstyle{definition}

\title{Quantify the Causes of Causal Emergence: Critical Conditions of Uncertainty and Asymmetry in Causal Structure}

\author[2,3,4,6]{Liye Jia}
\author[2,3,4,6]{Fengyufan Yang}
\author[3]{Ka Lok Man}
\author[3]{Erick Purwanto}
\author[3]{Sheng-Uei Guan}
\author[4]{Jeremy Smith}
\author[1,2,6 *]{Yutao Yue}

\affil[1]{Thrust of Artificial Intelligence and Thrust of Intelligent Transportation, The Hong Kong University of Science and Technology (Guangzhou), Guangzhou 511400, China}
\affil[2]{Institute of Deep Perception Technology, JITRI, Wuxi 214000, China}
\affil[3]{School of Advanced Technology, Xi’an Jiaotong-Liverpool University, Suzhou 215123, China}
\affil[4]{Department of Electrical Engineering and Electronics, University of Liverpool, Liverpool L69 3BX, United Kingdom}
\affil[5]{Department of Mathematical Sciences, University of Liverpool, Liverpool L69 3BX, United Kingdom}
\affil[6]{XJTLU-JITRI Academy of Industrial Technology, Xi‘an Jiaotong-Liverpool University, Suzhou 215123, China}
\affil[*]{Correspondence: \texttt{yutaoyue@hkust-gz.edu.cn}}

\begin{document}
\maketitle
\begin{abstract}
Beneficial to advanced computing devices, models with massive parameters are increasingly employed to extract more information to enhance the precision in describing and predicting the patterns of objective systems. This phenomenon is particularly pronounced in research domains associated with deep learning. However, investigations of causal relationships based on statistical and informational theories have posed an interesting and valuable challenge to large-scale models in the recent decade. Macroscopic models with fewer parameters can outperform their microscopic counterparts with more parameters in effectively representing the system. This valuable situation is called "Causal Emergence." This paper introduces a quantification framework, according to the Effective Information and Transition Probability Matrix, for assessing numerical conditions of Causal Emergence as theoretical constraints of its occurrence. Specifically, our results quantitatively prove the cause of Causal Emergence. By a particular coarse-graining strategy, optimizing uncertainty and asymmetry within the model's causal structure is significantly more influential than losing maximum information due to variations in model scales. Moreover, by delving into the potential exhibited by Partial Information Decomposition and Deep Learning networks in the study of Causal Emergence, we discuss potential application scenarios where our quantification framework could play a role in future investigations of Causal Emergence.
\end{abstract}


\thispagestyle{plain}
\setcounter{page}{1}
\section{Introduction}
\label{sec-intro}
We create many informative representations to interpret and manipulate the observed systems in the real world. The informational representation is often in the form of concepts (e.g., natural language words) or mathematical variables and their values (e.g., all forms of data) \cite{lloyd2002computational, yue2022world}. These indicators allow an accurate description of intuitive and abstract events to summarize their effects on the systems. Causality is a popular expression to describe the intrinsic relationship between an observed event and its impact \cite{paul2013causation,pearl2009causal, pearl2018book}. 

Surprisingly, simple mathematical models can describe many complex systems, such as those from quantum mechanics or macroeconomics. The simplicity of these mathematical models sometimes is the result of approximations (e.g., marginal revenue $\Delta MR = \frac{\Delta TR}{\Delta Q}$ in economics \cite{karl2019principles}), while sometimes seems to be intrinsic to the system (e.g., a photon energy $E=\hbar \nu$ in physics \cite{einstein1989uber}). These mathematical models represent the system's underlying causality, e.g., if a physicist changes the value of frequency $\nu$, the photon energy measurement will change accordingly. Hence, the causal relationships of complex systems have long been discussed to study objective regularities by various disciplines, such as philosophy, physics, and mathematics \cite{pepper1926emergence, bar2004mathematical, anderson1972more, klein2020emergence, baysan2020causal, pearl2000models, kivelson2016defining, fromm2005types}. However, although mathematical models can represent causality directly and intuitively, the derivation of math equations relies on ensuring a clear depiction of causal relationships within the system by analyzing the results of randomized controlled experiments based on professional knowledge from domain experts \cite{montgomery2017design, williams2013model, mead2017statistical}.

Therefore, in the current decade, people have begun to straightforwardly analyze the underlying causal relationships of the system from observational data or interventional results by using informational or statistical theories \cite{hoel2013quantifying, hoel2017map, chvykov2020causal, rosas2020reconciling, yao2021survey, zenil2017low} or applying human intervention to models, such as graphical models \cite{pearl2018book, gong2023causal} or biological networks \cite{zenil2019algorithmic}. When employing these methodologies, the experimenter's observation, specifically the information algorithms' preconditions and intervention experiments' designs, is pivotal for the system's causality investigations. In the context of information-theoretic methods, assumptions about the distribution of observed data are paramount, given that disparate data distributions yield entirely different computational results \cite{hoel2013quantifying, yao2021survey, balduzzi2011information}. For model-based approaches, the preliminary challenge lies in designing experiments that independently intervene in the system to observe the feedback \cite{pearl2018book, gong2023causal, zenil2019algorithmic}. Addressing these concerns is fundamental to the robust exploration of causal relationships.

On the other hand, the granularity of studies is another crucial factor in discovering causal relationships \cite{yue2022world}. For instance, neuroscience scientists can model the system from the microcolumns (considered the most microscopic granularity) \cite{buxhoeveden2002minicolumn}, the regions \cite{kaas1989does}, the cortex \cite{eliasmith2013build}, and the entire brain (regarded as the most macroscopic granularity) \cite{andersen2007hippocampal} to study the operational mechanisms of the brain from various levels. The models with different granularities reveal an interesting and valuable phenomenon. Macroscopic models, encompassing fewer parameters and systemic information, can better represent the system mechanism than microscopic models with more parameters containing additional and complete information about the system, as the phenomenon termed Causal Emergence (CE) by Hoel et al. \cite{hoel2013quantifying, hoel2017map, comolatti2022causal}.

Hoel and coworkers established a mathematical definition of CE using Effective Information (EI) by assuming that the model's interventions follow Maximum Entropy Distribution (MED) \cite{pearl2018book, hoel2013quantifying, comolatti2022causal}. The existence of CE has been repeatedly demonstrated and quantified in bits of the superior effectiveness of macroscopic models over microscopic counterparts by using artificially defined discrete \cite{hoel2017map} and continuous \cite{chvykov2020causal} system models as toy examples. Furthermore, recent research indicated that CE persists even when alternative metrics are employed \cite{comolatti2022causal} or when relaxing MED by different informational concepts \cite{rosas2020reconciling}. These discoveries have garnered widespread attention to the CE investigations. For example, Zhang and Liu \cite{zhang2022neural} devised a specific Machine Learning (ML) algorithm, offering a more efficient CE discovery than choosing coarse-graining strategies empirically. Moreover, Morrow, cooperating with Michaud and Hoel \cite{marrow2020examining}, has employed CE metrics, such as EI, as evaluation tools to interpret complex Deep Learning (DL) networks from a causality perspective. These efforts stimulate further explorations of CE's future developments and potential applications.

Nevertheless, existing research on CE mainly focuses on validating the CE induced by specific coarse-graining strategies or improving the design of coarse-graining strategies for deriving CE, with a minor emphasis on explaining the exact reasons behind why coarse-graining can lead to CE and exploring the particular conditions under which the strategies are effective for obtaining CE \cite{hoel2017map, chvykov2020causal, rosas2020reconciling, zhang2022neural}. In this work, we define the model's intrinsic characteristics, i.e., uncertainty and asymmetry, as factors affecting the effectiveness of discrete models based on Hoel's CE study \cite{hoel2017map}. Moreover, we introduce a quantification framework to compute the specific characteristics required by microscopic and macroscopic models while the CE occurs. The framework offers a numerical understanding of the reasons for the effectiveness of coarse-graining strategies in discovering CE. We aim to provide theoretical benchmarks as mathematical constraints for future CE research and offer valuable assistance for the extensive developments of CE theory.

In this paper, Section \ref{relate-works} provides an overview of prior CE research, establishing the background knowledge for our work. Simultaneously, we describe our study's scope and explain the choices of our research methods in this section. Section \ref{methods} explains our quantitative framework, detailing how synthetic models with controllable uncertainty and asymmetry are used by adjustive parameters to identify the sufficient and necessary conditions for CE occurrence. Section \ref{results_uncertainty} demonstrates why coarse-graining can derive the CE from models with significant uncertainty. Section \ref{results_asymmetry} validates the importance of reducing asymmetry for coarse-graining strategies to bring CE. Section \ref{discussion} discusses the assistance of our quantitative framework for CE research and potential improvements in the future based on additional experimental results. Finally, Section \ref{conclusion} summarizes the contributions of our work and its potential application scenarios for future research.

\section{Information-based Causal Emergency Theories}
\label{relate-works}
Causality, widely referenced across diverse disciplines, has various definitions \cite{pearl2018book, paul2013causation} and disparate research methodologies based on different academic domains \cite{yao2021survey, gong2023causal, zenil2019algorithmic, scholkopf2021toward}. In this paper, we aim to explore CE's critical conditions, i.e., the reduction of uncertainty and asymmetry by coarse-graining the models. Therefore, the statistical metrics of causal relationships are our primary concentration, particularly the work by Hoel \cite{hoel2017map} that aligns the EI computation with the model's numerical properties, the determinism and the degeneracy, to discover the CE occurrence. 

Hence, this section briefly introduces Hoel's framework for studying CE by the EI algorithm \cite{hoel2017map} in subsection \ref{hoel emergence}. Subsequently, in Section \ref{discuss_scope}, we discuss the primary reasons for selecting Hoel's work as our research background.
\subsection{Effective Information (EI), Transition Probabilistic Matrix (TPM), and Causal Emergence (CE)}
\label{hoel emergence}
EI is a metric that quantifies the influence of a specific intervention of the system's inputs $X$ on the distribution of the outputs $Y$ \cite{balduzzi2011information}. It utilizes the Kullback-Leibler divergence, $D_{KL}$ \cite{edwards2008elements}, to measure the variation of $Y$'s distribution when the intervention makes conditional variable X equal a specific value, $x_i$, where the intervention is denoted by Pearl's operator as $do(X=x_i)$ \cite{pearl2018book}. For instance, in a system with four possible states, denoted as $S = {00,\ 01,\ 10,\ 11}$, its inputs $X$ and outputs $Y$ can be represented by $X = S_t$ and $Y = S_{t+1}$, respectively. In this case, an artificial intervention on X, such as $do(X=00)$, will alter the distribution of Y, denoted as $P(Y=S_{t+1}|do(X=00))$, inducing a gap between the intervention distribution and pre-intervention distribution, indicated by the MED \cite{bishop2006pattern}, denoted as $H^{max}$. Then, EI can be a metric for quantifying the effect of the specific intervention, $do(X=00)$, on the system's outputs $Y$ by measuring the distance between the two distributions, as demonstrated by Equation \ref{EI_example} below \cite{hoel2013quantifying}.
\begin{align}
    \label{EI_example}
    EI = P(S_{t+1}|do(S_t=00)) \cdot \frac{P(S_{t+1}|do(S_t=00))}{H^{max}}
\end{align}

However, interventions are applied to the complex system's models \cite{pearl2009causal}, such as Markov Chains and TPMs \cite{robert2004markov}, graphical models \cite{gong2023causal}, gene models \cite{zenil2019algorithmic}, ML, or DL models \cite{scholkopf2021toward}. The quality of the models \cite{karnopp2012system}, instead of the interventions, is another influence on our study of inherent causal relationships within the system. Hoel et al. \cite{hoel2013quantifying, hoel2017map} refined the EI causal metric to assess whether models effectively represent causality within complex systems.

\begin{figure}[t] 
\centering 
\includegraphics[scale=0.38]{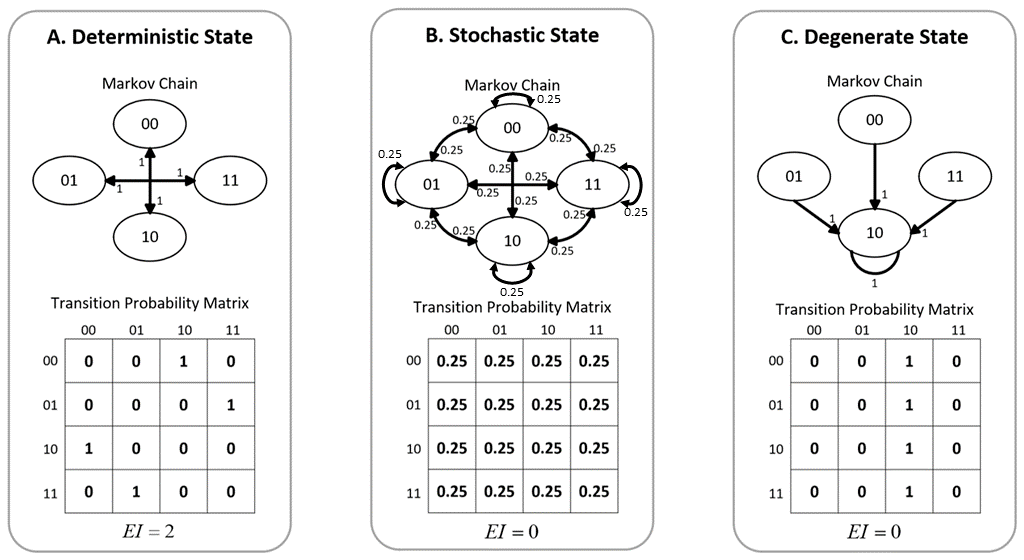}  
\caption{Deterministic, Stochastic, and Degenerate State Dynamics and TPMs of Three Systems. Both State Dynamics and TPMs show that mechanism of system to transfer current states $S_t=\{00,\ 01,\ 10,\ 11\}$ into future states $S^F$ following conditional probabilities in every row of TPMs. While the distribution of current states satisfies the MED, TPMs become numerical representations of causal structures of systems or models.} 
\label{TPMs_example} 
\end{figure}

Firstly, to demonstrate the model's effectiveness in representing causal relationships within the system, Hoel \cite{hoel2013quantifying} established an assumption to let the EI metric avoid the influence of prior knowledge embedded in the distribution of all the model's possible interventions, denoted as $I_D$. The assumption is that $I_D$ satisfies the MED principle, expressed by $I_D = H^{max}$. For instance, for the system with four states, as shown in Figure \ref{TPMs_example}, its models' potential interventions can be one of the current states $S_t=\{00,\ 01,\ 10,\ 11\}$. Under Hoel's assumption, $I_D=S_t = H^{max}$. Therefore, the probability of selecting each current state $s^c_t$ as an intervention $do(S_c=s^c_t)$ on the model is inversely proportional to the total number of current states, $N$, expressed as $P(do(S_t=s^c_t))=\frac{1}{N}$. In this example, the $I_D=\{\frac{1}{4},\ \frac{1}{4},\ \frac{1}{4},\ \frac{1}{4}\}$.

Consequently, the EI calculation can be related to the model's TPM \cite{hoel2017map}. Given the specific intervention as $do(S_c=s^c_t)$, the probability distribution of the model's outputs $Y$ is the conditional distribution of the future states, $S_{t+1}$, as denoted by $P(Y=S_{t+1}|do(S_t=s^c_t))$. Additionally, an average distribution, $E_D=\frac{\sum\limits^{N}_{c=1} P(Y=S_{t+1}|do(S_t=s^c_t))}{N}$, can indicate the model's average effect in transforming various interventions into their relative outputs. For example, as displayed by the stochastic state model in Figure \ref{TPMs_example}B, the distribution of the model's future states at the next time step, namely the probability distribution of the outputs, under the influence of one intervention, such as $do(S_t=00)$, is $P(S_{t+1}|do(S_t=00))=\{0.25,\ 0.25,\ 0.25,\ 0.25\}$. The model's average transformation effect is $E_D=\{\frac{0.25*4}{4},\ \frac{0.25*4}{4},\ \frac{0.25*4}{4},\ \frac{0.25*4}{4}\}=\{0.25,\ 0.25,\ 0.25,\ 0.25\}$.

These definitions enable two expressions \cite{hoel2017map}, $D_{KL}(p(S_F|do(S_C=s_c))||H^{max})$ and $D_{KL}(E_D|I_D)$. The first $D_{KL}$ operator measures the difference between the distributions of the model's outputs with and without a particular intervention (which distribution without intervening is the $H^{max}$. The second expression quantifies the model's overall effectiveness on the transformation from interventions to outputs (i.e., the total effects of interventions).

Furthermore, by normalizing the two $D_{KL}$ operators, Hoel \cite{hoel2017map} defined two intrinsic coefficients, $determinism$ and $degeneracy$, determining the model's effectiveness in representing the system. Equations \ref{determinism} and \ref{degeneracy} show the original algorithms \cite{hoel2017map} of the coefficients and our modified expressions, incorporating the distributions shown in each TPM's row, denoted as $row_i$, to provide a more intuitive process for computing $determinism$ and $degeneracy$.

\begin{align}
    \label{determinism}
    determinism &= \frac{1}{N}\sum\limits_{do(s_c)\in I_D}\frac{D_{KL}(p(S_F|do(S_C=s_c))||H^{max})}{\log_2(N)}\nonumber\\
                &=\frac{1}{N}\sum\limits_{i=1}^{N}\frac{D_{KL}(row_i||H^{max})}{\log_2(N)},\ \text{where } row_i \in TPM
\end{align}

\begin{align}
    \label{degeneracy}
    degeneracy=\frac{D_{KL}(E_D||I_D)}{\log_2(N)}&=\frac{D_{KL}\left(\sum\limits^N_{i=1}row_i/N||I_D\right)}{\log_2(N)},\nonumber\\ 
              &\ \ \ \ \ \ \ \ \text{where } row_i \in TPM
\end{align}

Finally, Hoel \cite{hoel2017map} defined the coefficient of a CM (CM) with $N$ states, $eff(CM)$, as the normalized effectiveness in representing causal relationships within the complex system. However, the CM can receive more system information while increasing the number of states \cite{grunwald2007minimum}. Therefore, the potential maximum effectiveness of a CM in describing causal relationships can be expressed by the $I_D$'s entropy \cite{mackay2003information}, indicated as $H(I_D)$. Combining the $eff(CM)$ with the $H(I_D)$, Equation \ref{EI_decop} provides the EI calculation for measuring the causal capacity of CM. On the other hand, Equation \ref{effectiveness} demonstrates how to compute the $eff(CM)$ through the CM's $determinism(CM)$ and $degeneracy(CM)$, reflecting the relationship between the EI of CM and its intrinsic coefficients.

\begin{align}
    \label{EI_decop}
    EI = H(I_D(\text{CM})) * eff(\text{CM})
\end{align}

\begin{align}
    \label{effectiveness}
    eff(\text{CM}) = determinism(\text{CM}) - degeneracy(\text{CM})
\end{align}

\begin{figure}[b] 
\centering 
\includegraphics[scale=0.4]{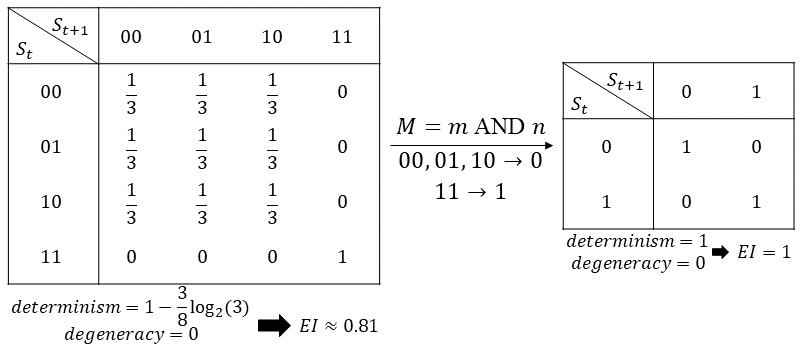} 
\caption{The example of CE while coarse-graining a CM\_m with four states into a CM\_M with two states \cite{hoel2017map}.} 
\label{CE_example} 
\end{figure}

According to the definitions above, the macroscopic CM (CM\_M) with fewer states should have a lower ability to represent causal relationships within the system than the microscopic CM (CM\_m) with more states. This expectation arises because the CM\_M's maximum causal capacity, $H(I_D(\text{CM\_M}))$, must be smaller than that of CM\_m. Interestingly, in the experiment \cite{hoel2017map}, the EI of CM\_M can sometimes exceed that of CM\_m. As an example illustrated in Figure \ref{CE_example}, a CM\_m with four states has a representational capacity whose maximization is $H(I_D(\text{CM\_m}))=2$, which is greater than that of a CM\_M with two states, which is $H(I_D(\text{CM\_m}))=1$. However, the value of CM\_m's EI (0.81 bits) is less than that of the CM\_M (1 bits). This phenomenon, where CM\_M exhibits greater causal capacity than CM\_m, is named "Causal Emergence" (CE) by Hoel \cite{hoel2013quantifying, hoel2017map, chvykov2020causal, comolatti2022causal}.

Through extensive analysis of CE experiments \cite{hoel2013quantifying, hoel2017map} and comparing this phenomenon with Shannon's information channel \cite{hoel2017map, cover1999elements}, Hoel preliminarily defined the causes of CE. Taking the examples of CM\_m and CM\_M presented in Figure \ref{CE_example}, a coarse-graining strategy by an AND logical operation applied to CM\_m merges the microscopic variables, $m$ and $n$, into the macroscopic variable $M$ of CM\_M. This coarse-graining strategy maps the microscopic states $00$, $01$, and $10$ to the macroscopic state $0$, while the remaining microscopic state $11$ maps to the macroscopic state $1$. In this process, the uncertainty in CM\_m, represented by the model's output probability distribution in the TPM as $\{\frac{1}{3},\ \frac{1}{3},\ \frac{1}{3},\ 0\}$, is eliminated, leading to an increase in CM\_M's $determinism$. Since the original CM\_m's $degeneracy$  is $0$, the coarse-graining strategy enhances the effectiveness coefficient to make that $eff(\text{CM\_M})>eff(\text{CM\_m})$, exceeding the difference in $H(I_D)$ between CM\_M and CM\_m. This change counterintuitively results in a higher EI value of CM\_M than that of CM\_m. Hoel indicated the CE occurrence when the change in EI by coarse-graining is positive, expressed as $\Delta EI = EI(CM_M) - EI(CM_m) > 0$ \cite{hoel2017map}. Furthermore, he generally explains the causes of CE as the gain in the effectiveness of models, denoted as $\frac{eff(CM_M)}{eff(CM_m)}$, being greater than the loss in maximum representation effectiveness, denoted as $\frac{size(CM_m)}{size(CM_M)}$. Hence, $\frac{eff(CM_M)}{eff(CM_m)}>\frac{size(CM_m)}{size(CM_M)}$ also serves as an indicator for CE caused by coarse-graining strategies \cite{hoel2017map}.

In modern modeling studies, scientists increasingly favor models with enormous parameters \cite{kenton2019bert,wu2023brief}, i.e., states, to describe the underlying regularities and causal relationships within complex systems as thoroughly as possible. The occurrence of CE provides theoretical support for the hypothesis that "small models can exhibit the same or even better performance compared to large models \cite{frankle2019mode}." Consequently, CE has attracted widespread attention. For example, Hoel and his collaborators continually expanded the CE framework \cite{chvykov2020causal} and demonstrated that CE is not solely dependent on EI as the metric \cite{comolatti2022causal}. Rosas et al. \cite{rosas2020reconciling} attempted to alleviate the limitations of the MED assumption on CE using the Partial Information Decomposition (PID) theory \cite{williams2010nonnegative}, contributing to further developments in CE research. Besides, Marrow and colleagues \cite{marrow2020examining} applied CE to the interpretability of DL networks, while Zhang and Liu \cite{zhang2022neural} used DL networks to discover CE. Similarly, we value the considerable role of CE in studying modeling approaches and optimizations. Therefore, we expect to propose more specific CE critical conditions than those described indicators to support future research in the CE direction.

\subsection{Our Scope and Reasons}
\label{discuss_scope}
To quantitatively specify more concrete CE critical conditions, i.e., the reduction of uncertainty and asymmetry in CM caused by coarse-graining strategies, our work is grounded in the theoretical framework of statistical and informational causality research \cite{yao2021survey}. Specifically, we adopt Hoel's classical framework that relies on EI and TPM \cite{hoel2017map} to theoretically measure the specific conditions of uncertainty and asymmetry that CMs should exhibit when CE occurs. In this subsection, we present the detailed reason for choosing Hoel's classical framework \cite{hoel2017map} as the foundation for our quantification framework.

Firstly, numerous outstanding works in the causality domain study the causal relationships embedded in complex systems from a more practical perspective \cite{zenil2019algorithmic, gong2023causal, scholkopf2021toward}. However, a concrete mathematical constraint can provide convenience for various research. For instance, optimization algorithms, crucial in DL \cite{goodfellow2016deep} and signal processing \cite{orfanidis1995introduction}, heavily rely on mathematical conditions to regulate optimization objectives and directions \cite{boyd2004convex}. The methods based on statistics or information theory can offer numerous causal metrics \cite{hoel2013quantifying, hoel2017map, chvykov2020causal, comolatti2022causal, yao2021survey} to quantify the strength or effectiveness of causal relationships, enabling the computation of specific numerical conditions of our interested CE phenomenon.

On the other hand, some existing studies highlight that statistical or informational causality research methods overly depend on assumptions \cite{zenil2017low, yao2021survey}, such as the MED used by Hoel \cite{hoel2013quantifying}. Some scientists argue that these assumptions have unreasonable aspects and limit the practical applications of those algorithms \cite{rosas2020reconciling}. However, for studying the causal representational capacity of CM, the MED assumption is necessary \cite{hoel2017map, marrow2020examining}. For focusing on the model itself, MED eliminates the influence of the prior distribution in $I_D$ on the values of statistical metrics (such as EI). This independence can objectively evaluate the CM's capability to represent causal relationships within the system.

Finally, in Hoel's classical CE framework based on EI and TPM \cite{hoel2017map}, the inherent features of CM are defined and quantified into two coefficients: $determinism$ and $degeneracy$. These two coefficients are also excellent metrics of our interests, the uncertainty and asymmetry, as the inherent characteristics of CM that influence the causal capacity. Equations \ref{determinism} and \ref{degeneracy} offer a significant convenience for our research. Rosas et al. \cite{rosas2020reconciling}, based on the PID \cite{williams2010nonnegative}, can decompose their CE measurement into multiple sub-items to quantify the intrinsic properties of the models, possibly serving as a background theory for quantifying critical conditions of CE. However, the complicated and extensive calculations in PID theory \cite{williams2010nonnegative, james2018unique} lead us to adopt Hoel's classical CE theory \cite{hoel2017map} as the background knowledge for this work.

\section{Quantifying the critical conditions of the CE: The framework}
\label{methods}
This section introduces our research methodology based on Hoel's classical CE theory \cite{hoel2017map} to clarify our quantification framework's designs and objectives. Hoel proposed two coefficients, $determinism$ and $degeneracy$, to measure the CM's effectiveness in representing the system's causal relationships by calculating EI from the model's TPM. According to extensive experiments and comparisons with Shannon's information channels, the CE is defined and indicated as $\Delta EI > 0$ or $\frac{eff(\text{CM\_M})}{eff(\text{CM\_m})} > \frac{size(\text{CM\_m})}{size(\text{CM\_m})}$ when coarse-graining the CM\_m into a CM\_M with fewer states but more effective causal representation. As the metrics of the CM's intrinsic characteristics, the uncertainty and asymmetry, the $determinism$ and $degeneracy$ can potentially measure the particular numerical conditions of microscopic and macroscopic CMs for whether coarse-graining strategies can cause the CE. Based on these coefficients and TPM's EI calculation (as shown in Equations \ref{determinism}, \ref{degeneracy}, \ref{EI_decop}, \ref{effectiveness}) \cite{hoel2017map}, we design our quantification framework, extending Hoel's classic CE framework to provide a more specific explanation of why the specific coarse-graining strategy can bring the occurrence of CE. By quantifying the conditions that coarse-graining strategies should meet, we expect to offer numerical and concrete constraints to facilitate CE research. 

In this section, we present our quantification framework's designs and relative derivations, giving their corresponding interpretations of being used. Since the background of our work is the causal measurements derived from statistical and informational theories, such as EI \cite{hoel2013quantifying} and information entropy \cite{cover1999elements}, Section \ref{hypotheses} introduces three assumptions as prerequisites for our framework's implementation. Section \ref{equation_CQE} demonstrates a Conditions' Quantification Equation (CQE) derived from regularities observed in synthetic TPMs generated by the framework. This equation is crucial for constructing our framework's significant component,  measuring the numerical uncertainty and asymmetry conditions that microscopic and macroscopic CMs should satisfy when CE occurs. Finally, the pseudo-codes for implementing the quantification framework and derivations of the corresponding equation tool are provided in Appendices \ref{algorithms_frame}, \ref{derivation_eqs}, and \ref{extensions}.

\subsection{The design of our framework: Three hypotheses}
\label{hypotheses}

\begin{figure}[b] 
\centering 
\includegraphics[scale=0.23]{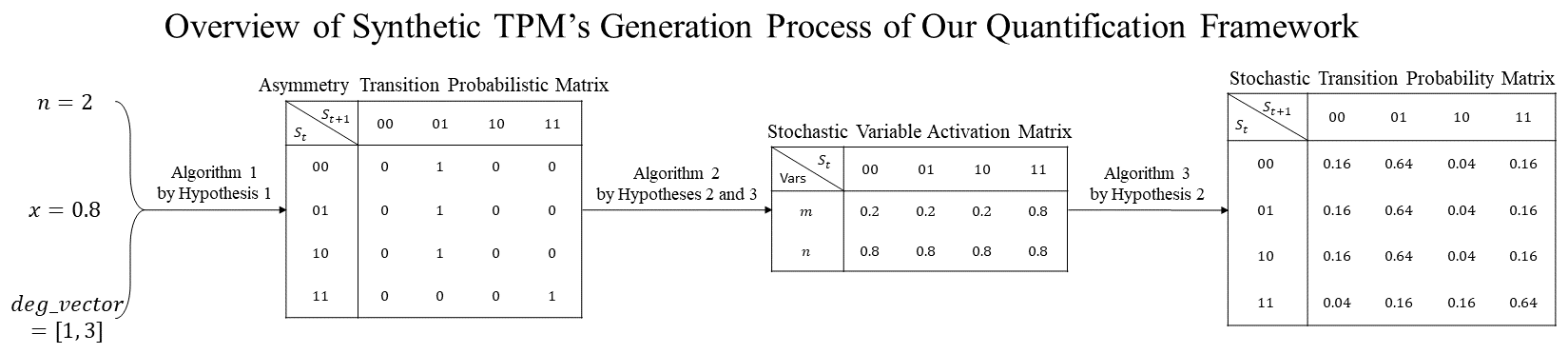}  
\caption{The overview of three hypotheses to construct three corresponding algorithms within Appendix \ref{algorithms_frame} as our framework's module for generating the synthetic TPMs to study the critical CE conditions.}
\label{overview1} 
\end{figure}

To assess the numerical conditions of uncertainty and asymmetry in CM for CE, we introduce two hyperparameters, $x$ and $deg\_vector$, into the quantification framework. The framework utilizes these parameters to generate a synthetic TPM of the CM with $N$ states, ensuring the desired uncertainty and asymmetry of the TPM for investigating critical characteristics when the CE occurs. Figure \ref{overview1} shows an overview of three hypotheses in this section to visualize their roles for our framework's TPM generation module. Here, $x$ controls the uncertainty in the synthetic TPM, while $deg\_vector$ governs the asymmetry. Setting two hyperparameters is motivated by the necessity to evaluate critical values of both CM's characteristics and separate the influences of uncertainty and asymmetry on the CE occurrence. Enhanced uncertainty diminishes $determinism$, while increased asymmetry augments $degeneracy$. These trends result in reduced CM's EI and then affect the CE.

\begin{figure}[t] 
\centering 
\includegraphics[scale=0.25]{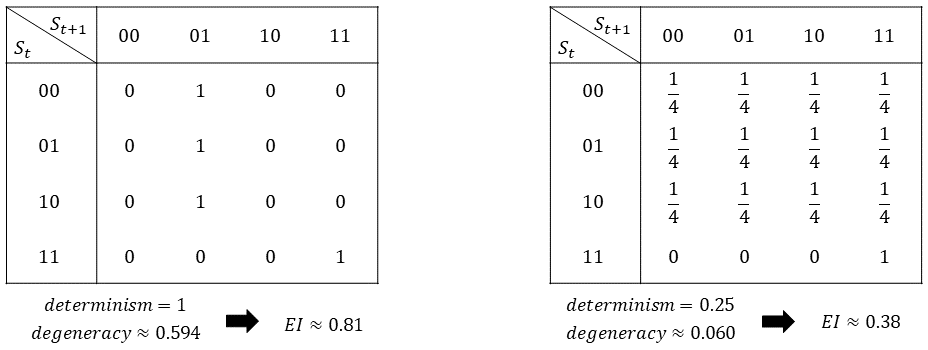}  
\caption{As the asymmetry determines the increment of the CM's $degeneracy$, this characteristic can independently influence the EI of the CM without any uncertainty. However, the values of the CM's asymmetry (measured by the $degeneracy$) can also be affected by the uncertainty (measured by the decrement of $determinism$).}
\label{x_influ_deg} 
\end{figure}

As shown in Figure \ref{x_influ_deg}, the left subfigure illustrates a deterministic but asymmetric TPM representing a CM with four states. Due to the asymmetry (quantified by the $degeneracy$ as approximately $0.594\text{ bits}$), the model's EI is approximately $0.81\text{ bits}$, instead of this model's maximum EI, denoted as $H(I_D) = 1\text{ bit}$. However, introducing uncertainty to three model states, $00$, $01$, and $10$, reduces the model's determinism from $1\text{ bit}$ to $0.25\text{ bits}$, which means there is $0.75\text{ bits}$ of uncertainty on the four-state CM, as shown in the right subfigure. Notably, the uncertainty simultaneously decreases the model's degeneracy to approximately $0.05\text{ bits}$, resulting in only a reduction of approximately $0.43\text{ bits}$ in EI (from $0.81\text{ bits}$ to $0.38\text{ bits}$), which is less than the decrease in determinism caused by uncertainty. Hence, our quantification framework requires separate control of uncertainty and asymmetry in generating synthetic TPMs using $x$ and $deg\_vector$, respectively, to isolate their effects on CM's EI and prevent unexpected mutual interactions. Additionally, in Figure \ref{x_influ_deg}, we observe that model uncertainty influences the asymmetry. Therefore, we consider asymmetry as an intrinsic characteristic represented by the CM. When generating the synthetic TPM, the first consideration of our framework is to implement the asymmetry specified by $deg\_vector$.

Building upon previous research on CE as outlined in works \cite{hoel2013quantifying, hoel2017map, comolatti2022causal}, it is understood that the TPM characterizes a model's asymmetry when multiple current states $S^{Red}_t$ redundantly transition to a single or multiple degenerate future states $s^fd_{t+1}\in S^{Deg}_{t+1}$. Therefore, we present our first hypothesis to derive the numerical representation of the TPM's asymmetry.

\begin{prop}
    \label{hypo_deg}
    An integer $FD$ and an array $CD$ can serve as numerical representations of a model's asymmetry. Specifically, $FD$ quantifies the amount of degenerate future states $S^{Deg}_{t+1}$ capable of receiving redundant information from multiple current states $S^{Red}_t$. Meanwhile, $FD$ corresponds to the $CD$'s length, denoted by Equation \ref{CD_len} below. 
    \begin{align}
        \label{CD_len}
        len(CD) = FD,
    \end{align}
    where $len(CD)$ represents the length of $CD$. This array contains the counts of redundant current states $s^c_t\in S^{Red}_t$. One number $cd_k$ in the $CD$ indicates how much current states redundantly transition to one degenerate future state. For a model featuring a total of $N$ states, $FD$ does not exceed $\frac{N}{2}$, and the sum of $CD$, denoted as $\sum CD$, does not surpass the total number of states, which is $N$.
\end{prop}

\begin{figure}[t] 
\centering 
\includegraphics[scale=0.25]{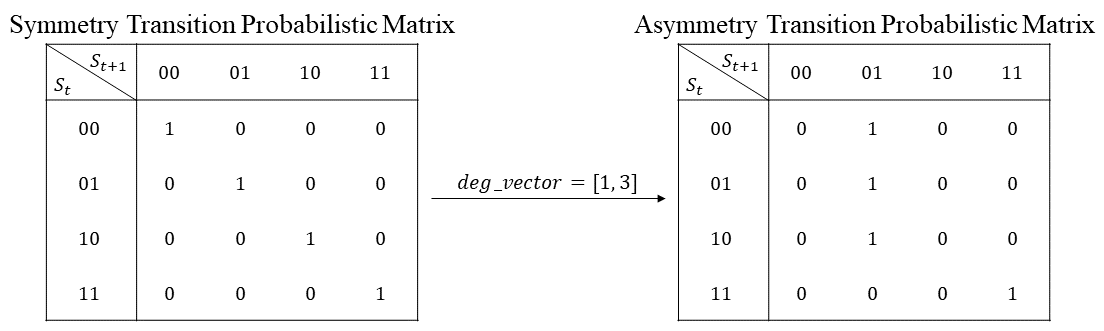}  
\caption{Implement the $deg\_vector=[1,\ 3]$ to transform a symmetric model into the asymmetric.} 
\label{deg_implemt} 
\end{figure}

To facilitate convenient control over the asymmetry, we introduce a binary array, the $deg\_vector$. This array's elements correspond to two critical parameters: the $FD$ and $\sum CD$, mentioned in Hypothesis \ref{hypo_deg}. The $deg\_vector$ provides a straightforward methodology for adjusting a model's asymmetry. For instance, to transform a symmetric TPM with four states into an asymmetric configuration, as shown by the left subfigure of Figure \ref{x_influ_deg}, Figure \ref{deg_implemt} shows the implementation process when setting the $deg\_vector = [1,\ 3]$. The $[1,\ 3]$ array means that the $FD=1$ and the $\sum CD = 3$ following the $deg\_vector$ definition. Hence, to generate an asymmetric TPM that fulfills the requirements, we designate the future state $01$ as the degenerate state and select the current states $00$, $01$, and $10$ as three redundant states. Subsequently, two rows of current states, $00$ and $10$, are replaced with the row of the current state $01$, as shown by the right subfigure of Figure \ref{deg_implemt}. The implementation of this process is elaborated in Algorithm \ref{implemt_alg_deg} within Appendix \ref{algorithms_frame}.

\begin{figure}[b] 
\centering 
\includegraphics[scale=0.25]{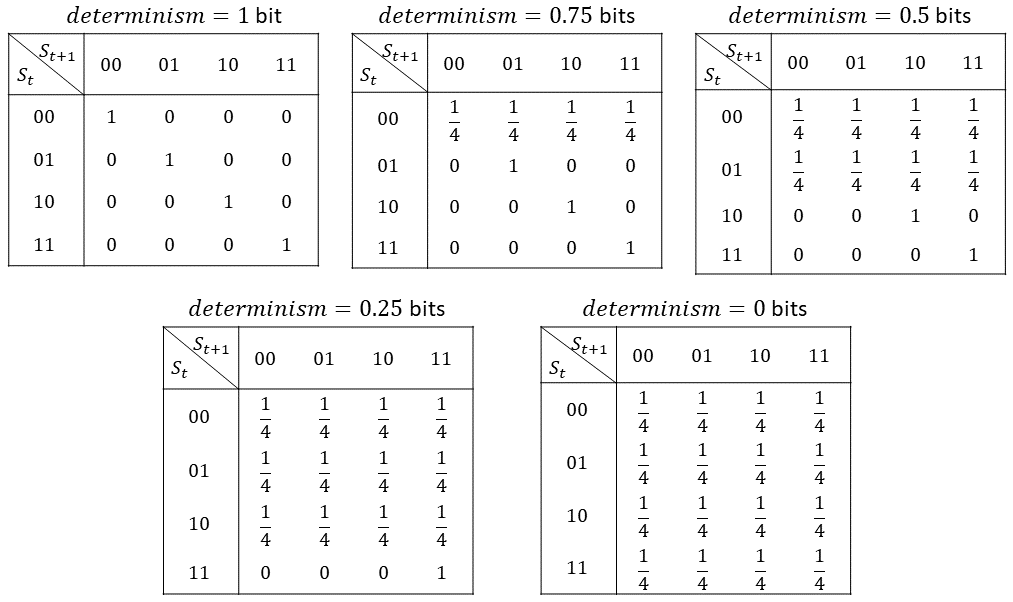}  
\caption{The $determinism$ will change discretely if the manipulation of the TPM's uncertainty is achieved by varying the state transition probabilities \cite{hoel2013quantifying,hoel2017map}.} 
\label{dis_det} 
\end{figure}

Another process related to our quantification framework's generation of synthetic TPMs requires a mechanism allowing us to control uncertainty in the TPM through a single variable $x$. Instead of discretely changing in $determinism$ by manipulating the TPM's state transition probabilities \cite{hoel2013quantifying,hoel2017map}, as displayed in Figure \ref{dis_det}, we aim to use the single variable $x$ to continuously and unidirectionally increase or decrease the generated TPM's $determinism$. For example, when the $x$ varies to introduce more uncertainty, $determinism$ decreases with the variation $\Delta x$ increases; conversely, if the variation $\Delta x$ indicates less uncertainty,  $determinism$ increases following changing the value of the $x$. This mechanism is more conducive to achieving our goal of identifying accurate numerical CE conditions as the constraints for coarse-graining strategies. However, substituting direct changes to state transition probabilities by a single variable $x$ raises a crucial question: How can the synthetic TPMs, whose uncertainty is controlled by $x$, satisfy the TPM definitions \cite{robert2004markov}?

Specifically, the synthetic TPMs generated by our quantification framework must satisfy two conditions  \cite{hoel2013quantifying, hoel2017map, robert2004markov}. Firstly, the sum of state transition probabilities in each row of the TPM equals 1, denoted as $\sum row_i = 1$. On the other hand, the average of state transition probabilities corresponding to all current states in the TPM equals 1, indicated as $\frac{\sum\limits^{N}_{i=1}\sum row_i}{N}=1$. To meet these constraints, we propose Hypotheses \ref{hypo_state2variable} and \ref{hypo_variable}. To thoroughly explain these two prerequisites, we sequentially describe the backgrounds of the mechanisms provided by the hypotheses and offer examples to illustrate their corresponding roles.

\begin{figure}[t] 
\centering 
\includegraphics[scale=0.25]{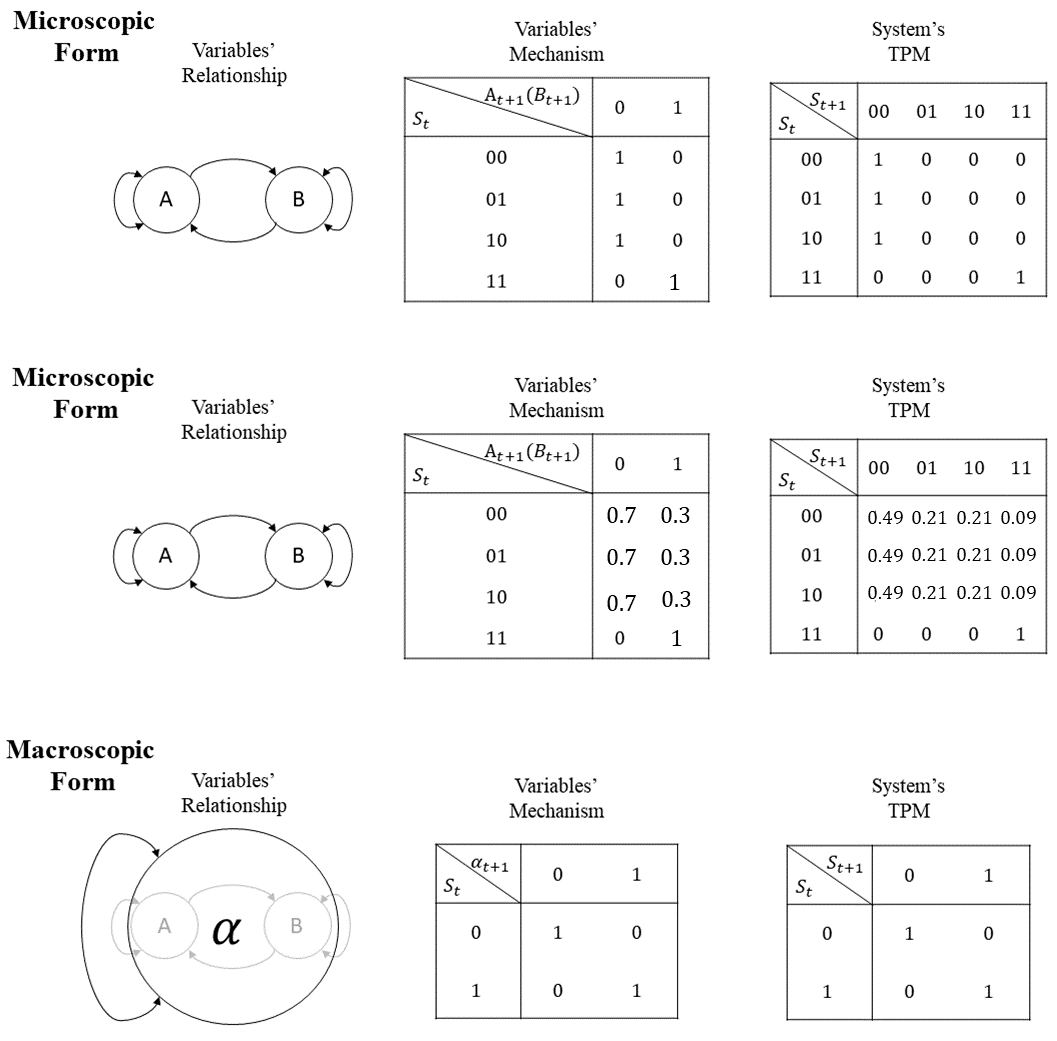}  
\caption{The Variables' Relationship, Variables' Mechanism, and TPM of the system's microscopic CMs and the macroscopic CM.} 
\label{logic} 
\end{figure}

Nevertheless, our primary concern is how to enable the parameter $x$ of our framework to control synthetic TPM's uncertainty to modify the value of $determinism$ continuously. Figure \ref{dis_det} has suggested that substituting state transition probabilities with the $x$ of TPMs is impractical. The discrete $determinism$ might lead to missing the specific numerical CE conditions on the CM's uncertainty. Moreover, it is too complex to satisfy the TPM's definitions by replacing the matrix elements with a single $x$. Therefore, we require a method to establish an indirect connection between $x$ and the state transition probabilities in synthetic TPMs. Fortunately, the approach of using the model's variable relationships to represent CMs proposed by Hoel and Tononi in their work inspires us. Figure \ref{logic} presents three graphical CM representations: Variables' Relationship, Variables' Mechanism, and TPM. For Microscopic Form at the top of Figure \ref{logic}, the logic of updating the variable value to the next time step can be explained by the logical AND operation, as demonstrated by the $A_{t+1}=A_t\text{ AND }B_t$. This relationship indicates both variables will be zero in the future until the current state is intervened into $11$, as shown by the Variables' Mechanism within Microscopic Form at the top of Figure \ref{logic}. 

Finally, the mechanism can derive the corresponding TPM as the representation of an asymmetric CM. These representations and their relations explain the uncertainty of the CM caused by the errors of logical operations for updating the variables. For example, assuming the AND operation has errors when current states are $00$, $01$, and $10$ leads to the variable's value at the next time step becomes probabilistic, such as the $P(A_{t+1}=0|S_t=00)=0.7$, as displayed by the Variables' Mechanism of Microscopic Form in the middle of Figure \ref{logic}. Consequently, this stochastic mechanism will derive an uncertain TPM representing a stochastic CM. Hence, we propose Hypothesis \ref{hypo_state2variable} based on Hoel and Tononi's approach to establish a process of transforming the deterministic TPM into the Variable Activation Matrix (VAM, which is similar to the Variables' Mechanisms in Figure \ref{logic}.) and reform the VAM back to the TPM as the synthetic representation of CM, whose uncertainty controlled by $x$ and asymmetry governed by $deg\_vector$, respectively.

\begin{prop}
    \label{hypo_state2variable}
        In the case of a discrete model comprising $N$ states, each of these states consists of $n$ binary variables. There exists a mathematical relationship between the $n$ and the $N$, which is expressed by Equation \ref{var2state} below:
        \begin{align}
        \label{var2state}
            N = 2^n \Rightarrow n = \log_2(N)
        \end{align}
        Supposing the transformation process from the Variables' Mechanism to the System's TPM in Figure 6 is reversible, we can derive a deterministic VAM from the deterministic TPM of CM. In this scenario, the conditional probability of the CM's variable being $1$ is determined by the future state $s^f_{t+1}$ to which corresponding current state $s^c_t$ transitions. Subsequently, given our assumption that this transformation is a reversible process, denoted as $TPM\Leftrightarrow VAM$, the VAM can reform back to the TPM as a comprehensive representation of the CM with controllable uncertainty and asymmetry.
\end{prop}

\begin{figure}[b] 
\centering 
\includegraphics[scale=0.25]{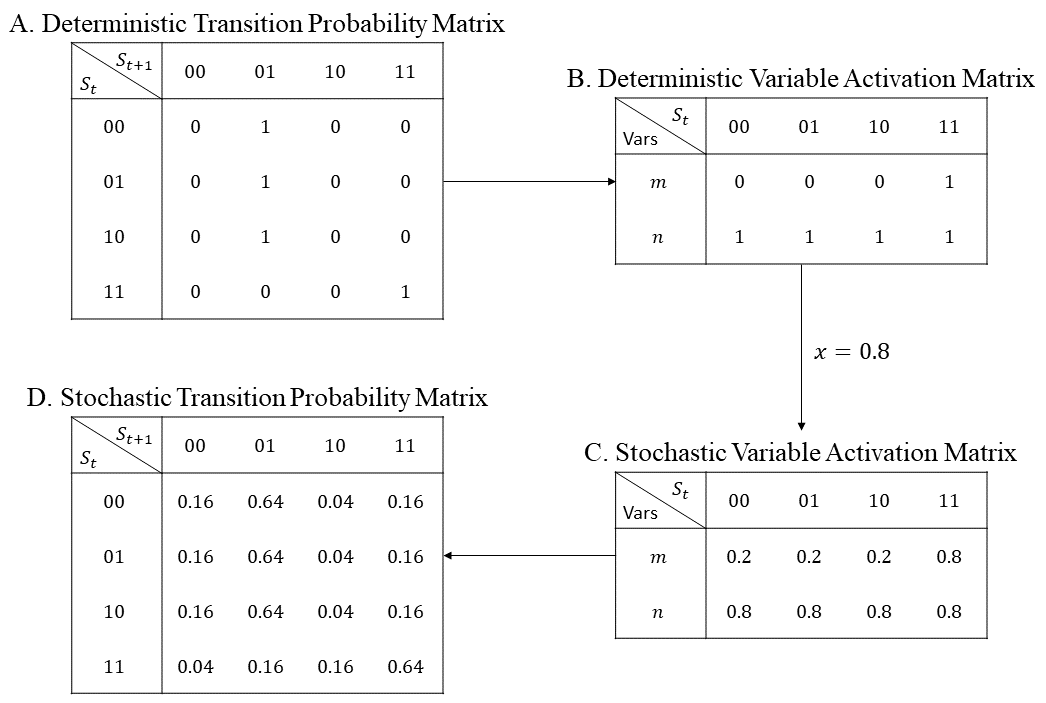}  
\caption{The process of applying the single parameter $x$ to control the model's uncertainty. The thorough process of transforming A into C is demonstrated by Algorithm \ref{trans_st2va} in Appendix \ref{algorithms_frame}.} 
\label{x_implemt} 
\end{figure}

Figure \ref{x_implemt} displays an illustrative example to explain Hypothesis \ref{hypo_state2variable} comprehensively. Consider an asymmetric CM with four states, characterized by $N=4$ and $n=2$, denoted as binary variables $m$ and $n$. The deterministic TPM, depicted in Figure \ref{x_implemt}A, reveals that the model's future state becomes $01$ given current states $00$, $01$, or $10$. In line with our hypothesis, this scenario implies that conditional probabilities of $m=1$ and $n=1$ in the next time step, given current states $00$, $01$, or $10$, are as follows: $P(m=1|do(S_t=00,\ 01,\ \text{or} 10))=0$ and $P(n=1|do(S_t=00,\ 01,\ \text{or} 10))=1$. Similarly, the conditional probabilities are $P(m=1|do(S_t=11))=1$ and $P(n=1|do(S_t=11))=1$, provided that the current state $11$ accurately transitions to the future state $11$. Consequently, we can derive the Deterministic VAM in Figure \ref{x_implemt}B from the deterministic TPM in Figure \ref{x_implemt}A.

Furthermore, given our assumption $TPM\Leftrightarrow VAM$, the VAM can be transformed back into a complete TPM to represent the CM whose uncertainty and asymmetry are controlled by our framework's two parameters, $x$ and $deg\_vector$, respectively. As illustrated in the transformation from Figure \ref{x_implemt}C to Figure \ref{x_implemt}D, introducing uncertainty that alters $P(m=1|S_t=00)$ and $P(n=1|S_t=00)$ of Deterministic VAM to $0.2$ and $0.8$ of Stochastic VAM. Since the CM's states are composed of combinations of the values of corresponding binary variables, such as $m$ and $n$ as this example, for the current state $00$, the conditional probabilities for its four possible future states can be computed as follows: $P(S_{t+1}=00|S_t=00) = P(m=0|S_t=00) \cdot P(n=0|S_t=00) = (1-0.2) \cdot (1-0.8) = 0.16$, $P(S_{t+1}=01|S_t=00) = P(m=0|S_t=00) \cdot P(n=1|S_t=00) = (1-0.2) \cdot 0.8 = 0.64$, $P(S_{t+1}=10|S_t=00) = P(m=1|S_t=00) \cdot P(n=0|S_t=00) = 0.2 \cdot (1-0.8) = 0.04$, and $P(S_{t+1}=11|S_t=00) = P(m=1|S_t=00) \cdot P(n=1|S_t=00) = 0.2 \cdot 0.8 = 0.16$. Following the same steps, we can calculate the conditional probabilities of transitions from the other three current states to each future state, obtaining the final TPM as the representation of CM. This synthetic TPM allows us to quantify the critical CE conditions by varying the settings of $x$ and $deg\_vector$ of our quantification framework.

In the mechanism proposed by Hypothesis \ref{hypo_state2variable}, the VAM allows us to establish an indirect connection between the $x$ and TPM's state transition probabilities by the conditional probability of the CM variable $v_i=1$ at the next time step. Due to the assumption $TPM\Leftrightarrow VAM$, this approach ensures that the synthetic TPM controlled by $x$ can satisfy two constraints of the TPM \cite{hoel2013quantifying,robert2004markov}. Specifically, we substitute the $v_i=1$ conditional probabilities, whose values are $1$ and $0$, in the Deterministic TPM with $x$ and $1-x$, respectively. As shown in the transformation from Figure \ref{x_implemt}B to Figure \ref{x_implemt}C, setting $x=0.8$ transforms Deterministic VAM into Stochastic VAM, introducing uncertainty controlled by a single parameter $x$ into the synthetic TPM generated by our quantification framework. To provide a more formalized explanation for this manipulation, we provide Hypothesis \ref{hypo_variable} as a clarification for this step.

\begin{prop}
    \label{hypo_variable}
    By transforming Deterministic TPM into Deterministic VAM, we can replace the conditional probabilities representing the CM variables equaling $1$ at the next time step with $x$ and $1-x$. This operation introduces uncertainty controlled by the single parameter $x$ into the synthetic TPM generated by our quantification framework, allowing manipulation of the TPM's $determinism$ to study CE critical conditions. Specifically, the relationship between the elements ($0$ or $1$) in the Deterministic VAM and $1-x$ or $x$ is defined by Equation \ref{x-1-x} below.
    \begin{align}
    \label{x-1-x}
    P(v_i=1|S=s^c_t) = 
        \begin{cases}
            1-x, & \text{ if }P(v_i=1|S=s^c_t)=0\text{ in the Deterministic VAM}\\
            x, & \text{ if }P(v_i=1|S=s^c_t)=1\text{ in the Deterministic VAM}
        \end{cases}
    \end{align}
\end{prop}

In summary, the implementation of Hypotheses \ref{hypo_state2variable} and \ref{hypo_variable} are provided by Algorithms \ref{trans_st2va} and \ref{trans_va2st} within Appendix \ref{algorithms_frame}, respectively. Combining these with Algorithm \ref{implemt_alg_deg} implemented based on Hypothesis \ref{hypo_deg}, three algorithms in Appendix \ref{algorithms_frame} constitute the modules of our quantification framework for generating synthetic TPM. The uncertainty and asymmetry of this TPM are controlled by the framework's parameters, $x$ and $deg\_vector$, enabling us to artificially manipulate the values of $determinism$ and $degeneracy$ for studying the specific numerical conditions of CE. Furthermore, the work of Hoel and Tononi \cite{hoel2013quantifying} demonstrates that logical operations can be coarse-graining strategies to merge microscopic variables into macroscopic variables, yielding a more effective macroscopic CM to represent causal relationships in the system. For example, the Macroscopic Form at the bottom of Figure \ref{logic} can be obtained by applying the AND logical operation to combine microscopic variables, $A$ and $B$, into the macroscopic variable $\alpha$, denoted as $\alpha = A \text{ AND } B$. This coarse-graining eliminates the asymmetry in the original microscopic TPM to enhance EI. These ideas inspire further applications of our quantification framework, which will be preliminarily validated and analyzed in Section \ref{discussion}.

\subsection{Relationship between $determinism$ and $x$ and Conditions' Quantification Equation (CQE)}
\label{equation_CQE}

Following the generation module described in subsection \ref{hypotheses} above, our quantification framework can output synthetic TPMs, whose uncertainty and asymmetry are governed by two parameters, $x$ and $deg\_vector$, for studying the numerical conditions of CE occurrence. This design allows our framework to discover precise values of $determinism$ and $degeneracy$ of CM as critical CE constraints, inducing the EI of CM\_m to be identical to that of CM\_M, where the CE indicator $\Delta EI = EI({\text{CM\_M}}) - EI({\text{CM\_m}}) >0$ will be achieved if uncertainty or asymmetry increases a bit. Figure \ref{quan_process1} illustrates this quantification process implemented by Algorithm \ref{TPM_solver} in Appendix \ref{extensions}.

\begin{figure}[t] 
\centering 
\includegraphics[scale=0.23]{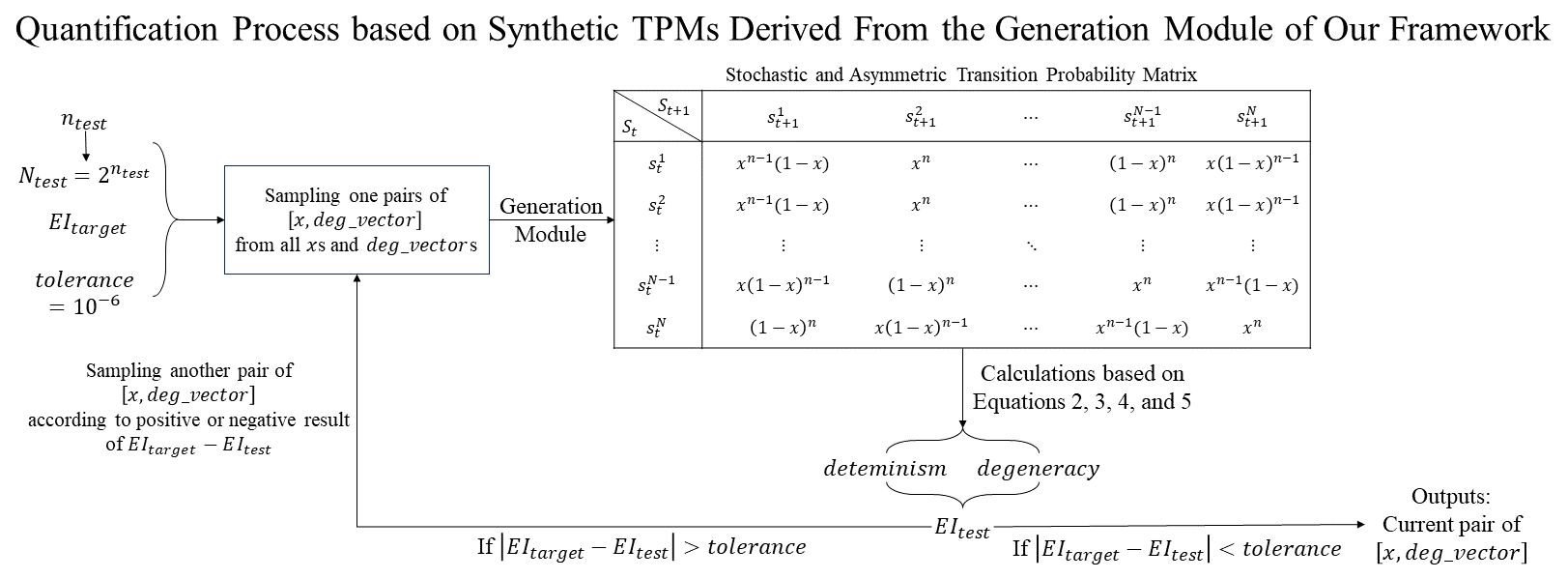}  
\caption{The quantification process based on synthetic TPMs derived from our framework's generation module in Figure \ref{overview1}. This process accepts the number of tested CM's variables $n_{test}$, the EI of targeted CM $EI_{target}$, and an error tolerance $tolerance$ as inputs. From the final synthetic TPM, the EI of tested CM with a sampled pair of $[x,\ deg\_vector]$ will be calculated by Equations \ref{determinism}, \ref{degeneracy}, \ref{EI_decop}, and \ref{effectiveness}. Finally, if the $|EI_{target}-EI_{test}|<tolerance$ satisfies, this process will output the current pair of $[x,\ deg\_vector]$ as critical CE conditions. Conversely, the module will choose another unrepeated pair of $[x,\ deg\_vector]$ for the next iteration. The thorough implementation of this process is provided by Algorithm \ref{TPM_solver} in Appendix \ref{extensions}.
}
\label{quan_process1} 
\end{figure}

However, the computational complexity of Figure \ref{quan_process1}'s quantification process will prevent efficient experiments and measurements of large CMs with many variables. Specifically, for calculating the $D_{KL}(row_i||H^{max})$ \cite{edwards2008elements}, Algorithm \ref{TPM_solver} requires actuating $N$ times of $\log_2(\cdot)$ operations and multiplications following $N-1$ times of additions. Hence, the computation of $determinism$ from the synthetic TPM is implemented by at least $N^2$ times of $\log_2(\cdot)$ operations and multiplications with $N\times (N-1)$ additions to combine the products. Conversely, the $degeneracy$ is easier to obtain through one time $D_{KL}\left(\sum\limits^N_{i=1}row_i/N||I_D\right)$ following $N$ times of $\log_2(\cdot)$ operations and multiplications and $N-1$ times of additions, since it is calculated by the average of each TPM's row, denoted as $\sum\limits^N_{i=1}row_i/N$. As CM's state number $N$ is exponentially increasing relative to the number of CM's variables $n$, the computational complexity of deriving the $EI_{test}$ from the synthetic TPM is exponentially enhancing when the $n_{test}$ is a large number, especially for the calculation of $determinism$. Therefore, to improve the efficiency of our quantification framework and validate the critical CE conditions with large CMs, we attempt to optimize the complexity of computing processes. 

\begin{figure}[b] 
\centering 
\includegraphics[scale=0.2]{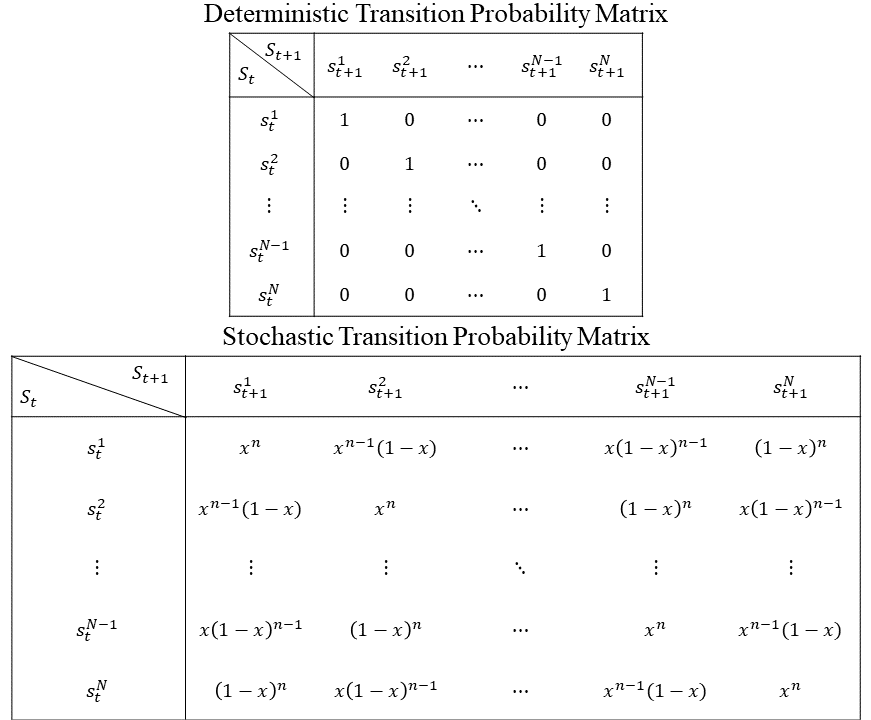}  
\caption{Transforming a symmetric and deterministic TPM with $n$ variables into stochastic format in
which conditional probabilities displayed by polynomials composed of $x$ and $n$.}
\label{x-n} 
\end{figure}

\begin{figure}[t] 
\centering 
\includegraphics[scale=0.32]{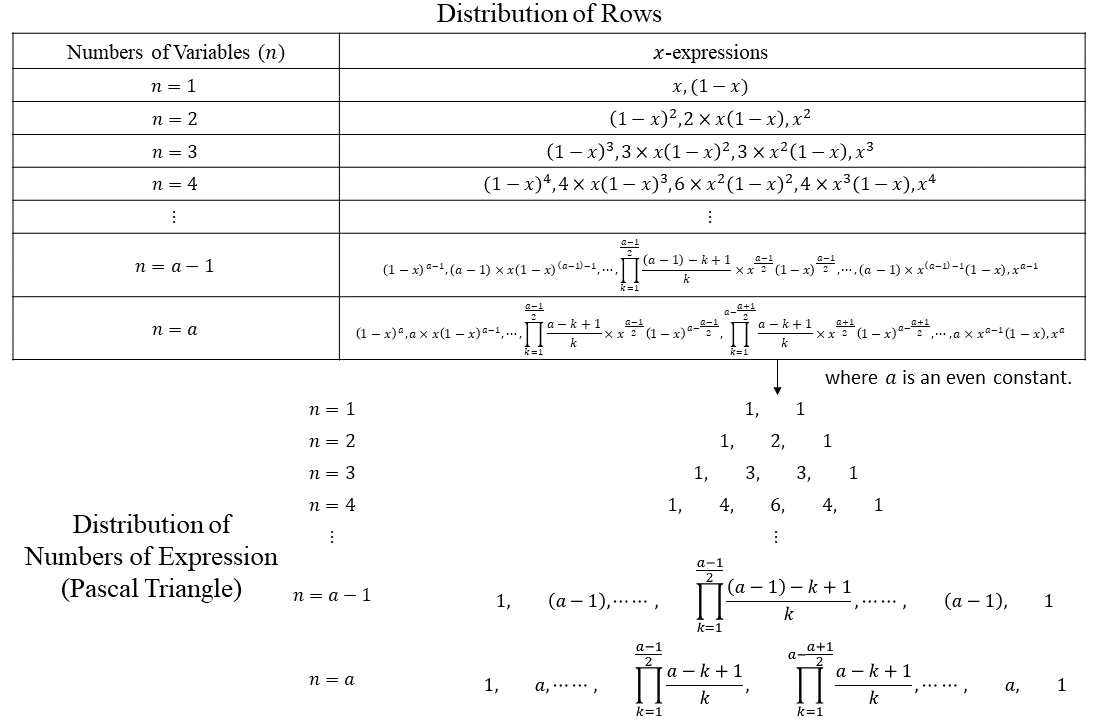}  
\caption{The regularity of $x$ and $n$ polynomials contained in TPM's rows relative to model's variable number $n$. The numbers of each polynomial in one row become a Pascal Triangle following the increment in the $n$.}
\label{regularity} 
\end{figure}

Fortunately, through experiments on various settings of $deg\_vector$ and $n$, we have discovered two regularities within the framework's synthetic TPMs. Specifically, each row of TPMs contains identical numbers of specific types of polynomials constructed by $x$ and $n$, as shown in Figure \ref{x-n}. The types of polynomials and their quantities in one row of TPM are determined by the CM's variable number $n$. "Distribution of Rows" Table in Figure \ref{regularity} shows four instances of regularities. For CMs with variable numbers of $n=1$ and $n=2$, row's distributions of their TPMs contain the polynomials ${x,\ 1-x}$ and ${(1-x)^2,\ x(1-x),\ x^2}$, respectively. In the case of the $n=1$, every TPM's row has one polynomial of $x$ and one polynomial of $1-x$, whereas each row of $n=2$ case contains one $(1-x)^2$, two $x(1-x)$, and one $x^2$.

Through the pattern of each row in the synthetic TPM containing the same types and numbers of $x$ polynomials, the calculation of $\sum\limits_{row_i\in TPM}D_{KL}(row_i||H^{max})$ can be modified into Equation \ref{dkl_new} below.

\begin{align}
    \label{dkl_new}
    \sum\limits_{row_i\in TPM}D_{KL}(row_i||H^{max}) = \left(\sum\limits_{p_i\in row_i}p_i\cdot \log_2\left(\frac{p_i}{\frac{1}{N}}\right)\right) * N
\end{align}

On the other hand, through another regularity that each row contains specific prototypes constructed by $x$ and $n$, we expect that Equation \ref{dkl_new} can be further simplified into a form of the expression of two parameters. By examining row distributions of synthetic TPMs with $n=1$, $n=2$, $n=3$, and $n=4$ number of variables, as presented in Figure \ref{regularity}, we have identified two general equations that allow us to derive all the kinds of polynomials and their quantities in every TPM's row directly from the $n$. For the types of polynomials present in each row, we can use Equation \ref{reg_polkinds}, which repeatedly adds $x^i(1-x)^{(n-i)}$ to the set with an increasing exponent $i\in [0,\ n]$. This process finally results in a group of polynomials representing the TPM's row distribution, such as $\{(1-x)^2,\ x(1-x),\ x^2\}$ of the synthetic $n=2$ TPM. Regarding the quantities of each polynomial kind, Pascal's Triangle \cite{rosen2007discrete} provides a valuable illustration of how the numbers of each polynomial type in every TPM's row vary with the variable number $n$. As a result, quantities of polynomials depend on the $n$ and the $i$, indicating the index of polynomial items. As demonstrated by Equation \ref{reg_polnums}, the amount of each kind is the result of iterative multiplications of $\frac{n-k+1}{k}$, where the integer $k$ starts from $1$ and ends at the value of polynomial item's index, which is $i$ or $n-i$.
\begin{align}
    \label{reg_polkinds}
    set = \{x^i(1-x)^{n-i}\}, \text{ where } i\in [0,\ n]
\end{align}
\begin{align}
    \label{reg_polnums}
    num = 
    \begin{cases}
        \prod\limits^i_k \frac{n-k+1}{k}, & \text{ if }i\leq (n-i)\\
        \prod\limits^{n-i}_k \frac{n-k+1}{k}, & \text{Otherwise}
    \end{cases}
\end{align}

Equations \ref{reg_polkinds} and \ref{reg_polnums} and Figure \ref{regularity}'s Pascal's Triangle prove the synthetic TPM's pattern. Each row distribution contains particular types of $x$ and $n$ polynomials, and each type's quantities depend on the parameter $n$. These results imply a straightforward expression of $x$ and $n$ for calculating the $D_{KL}(row_i||H^{max})$ \cite{edwards2008elements} and then simplifying the computation of $determinism$ to derive the $EI_{test}$ from synthetic TPMs more efficiently. Based on row distributions within the cases of $n=1$, $n=2$, and $n=3$, the $D_{KL}(row_i||H^{max})$ \cite{edwards2008elements} can be expressed using Equation \ref{reg_kl}. The detailed derivation of this equation based on models with three examples and its validation by the row distribution with $n=4$ are in Appendix \ref{derivation_eqs}.
\begin{align}
    \label{reg_kl}
    D_{KL}(row_i||H^{max})=n(1+(1-x)\log_2(1-x)+x\log_2(x))
\end{align}

\begin{figure}[b] 
\centering 
\includegraphics[scale=0.5]{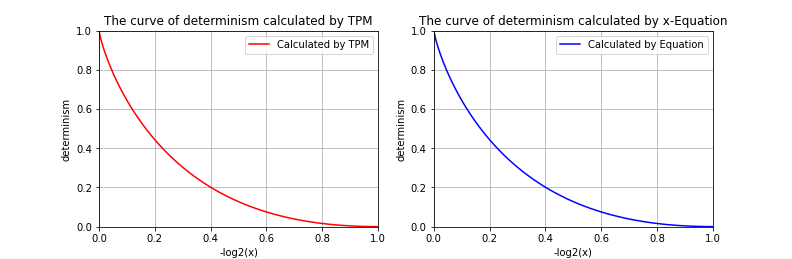}  
\caption{The comparison between the blue curve, representing $determinism$ calculated by Equation \ref{reg_det}, and the red curve represents $determinism$ computed from a $n=4$ scaled TPM generated by our framework. Two curves overlap perfectly, confirming that Equation \ref{reg_det} accurately captures the relationship between the parameter $x$ and model's $determinism$.}
\label{reg_det_verify} 
\end{figure}

Combining Equations \ref{dkl_new} and \ref{reg_kl} into Hoel's $determinism$ calculation (i.e., Equation \ref{determinism} \cite{hoel2017map}), we can obtain a simplified expression that allows the straightforward computation of $determinism$ of synthetic TPM from the parameter $x$. Equation \ref{reg_det} below describes the derivation and final expression.
\begin{align}
    \label{reg_det}
    determinism&=\frac{1}{N}\sum\limits_{i=1}^{N}\frac{D_{KL}(row_i||H^{max})}{\log_2(N)}\nonumber\\
               &=\frac{1}{N}*\frac{\sum\limits_{row_i\in TPM}D_{KL}(row_i||H^{max})}{\log_2(N)}\nonumber\\
               &=\frac{1}{N}*\frac{[n(1+(1-x)\log_2(1-x)+x\log_2(x))]*N}{n}\nonumber\\
               &=\frac{1}{N}*[1+(1-x)\log_2(1-x)+x\log_2(x)]*N\nonumber\\
               &=1+(1-x)\log_2(1-x)+x\log_2(x)
\end{align}

\begin{figure}[t] 
\centering 
\includegraphics[scale=0.23]{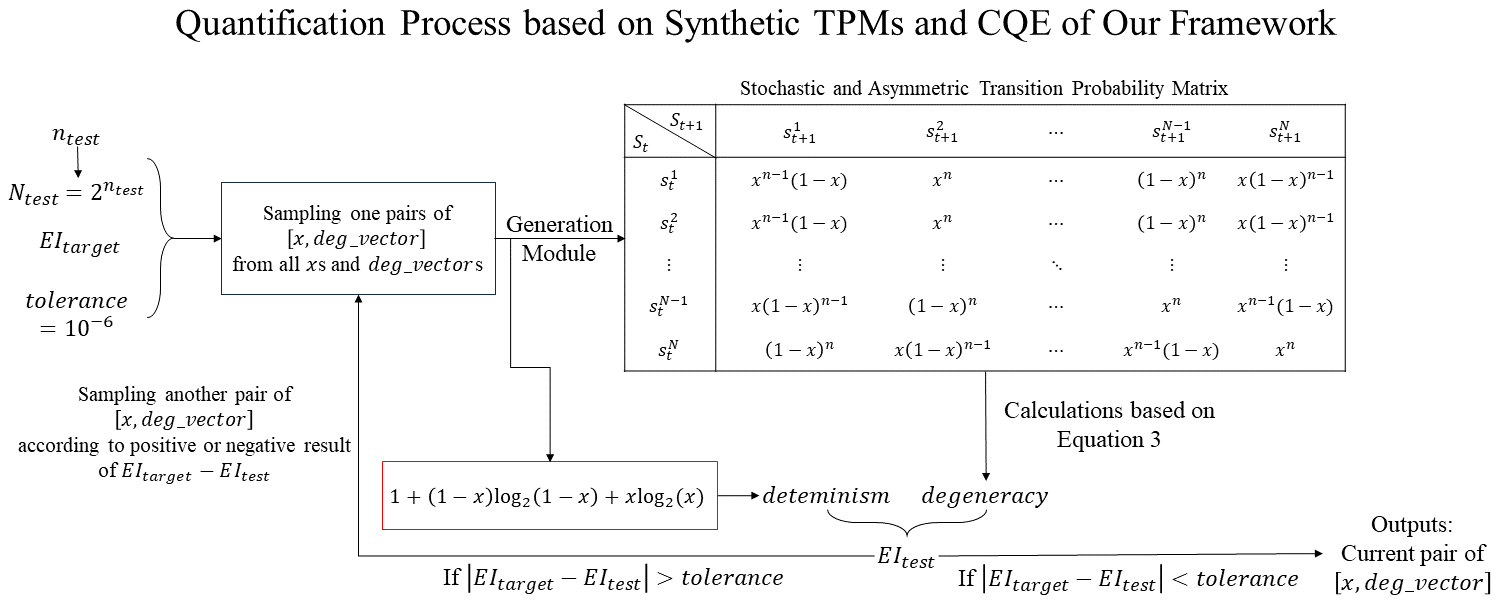}  
    \caption{The quantification process based on synthetic TPMs and CQE. By applying the simplified equation (circled by the red rectangle) to straightforwardly calculate CM's $determinism$ from the parameter $x$ in a pair of sample $[x,\ deg\_vector]$, we can substitute the calculations of $N^2$ times of $\log_2(\cdot)$ operations and the same times of multiplications with two times of those computations. Hence, this new quantification process based on CQE has less computational complexity than that of Figure \ref{quan_process1}. The implementation of this process is provided by Algorithm \ref{solver} in Appendix \ref{extensions}.}
\label{quan_process2} 
\end{figure}

Furthermore, Equation \ref{reg_det} demonstrates that our framework effectively controls the uncertainty of CMs through the parameter $x$. However, using $x$ values as numerical representations of uncertainty is impractical, as this parameter falls within the range of $[0.5,\ 1]$ to signify conditional probabilities of each model's variable $v_i=1$, defined by Hypothesis \ref{hypo_variable}. Figure \ref{reg_det_verify} compares continuous $determinism$ curves calculated by the synthetic TPM and Equation \ref{reg_det}, respectively, to validate the correctness of this simple expression. Here, we replace $x$ with $-\log_2(x)$, within the range $[0,\ 1]$, to more suitably represent the model's uncertainty. For example, $-\log_2(x) = 0$ denotes a deterministic CM, whose $EI=H(I_D)=n$, while $-\log_2(x) = 1$ represents a stochastic CM with $EI = 0$. Equation \ref{reg_uncer} provides a mathematical expression for our definition of uncertainty. We can derive a Conditions' Quantification Equation (CQE) by combining Equation \ref{reg_det} with Equation \ref{EI_decop},  as demonstrated by Equation \ref{reg_cqe}. The CQE provides an efficient quantification process for computing the critical CE conditions. Figure \ref{quan_process2} and Algorithm \ref{solver} within Appendix \ref{extensions} separately describe an overview and implementation of the new process.
\begin{align}
    \label{reg_uncer}
    uncertainty = -\log_2(x) \text{ bits}
\end{align}
\begin{align}
    \label{reg_cqe}
    1+(1-x)\log_2(1-x)+x\log_2(x) - degeneracy = \frac{EI}{n}
\end{align}

As shown in Figure \ref{quan_process2}, based on the CQE, the quantification process can replace complex $determinism$ computations with two times of $\log_2(\cdot)$ operations and that times of multiplications, as shown by the block circled by the red rectangle. However, the new quantification still requires synthetic TPMs to calculate the $degeneracy$ through Equation \ref{degeneracy}. Hence, the generation module of our framework described in subsection \ref{hypotheses} is executed once for each iteration to find critical CE conditions from the parameter spaces of $x$ and $deg\_vector$. This mechanism introduces additional operation complexity into our framework. In Section \ref{discussion}, we provide an extensive experiment for attempting to solve this limitation through neural networks trained by data that our quantification framework in Figure \ref{quan_process2} generates. Nevertheless, the current quantification process can calculate numerical conditions for CE when coarse-graining a CM\_m with $n=11$ variables into any of its CM\_Ms. Therefore, the experimental results shown in the rest of the contents in this article are derived from the framework illustrated in Figure \ref{quan_process2}.

\section{Quantifying the critical conditions of CE: The reduction of CMs' uncertainty}
\label{results_uncertainty}

\begin{figure}[t] 
    \centering 
    \includegraphics[scale=0.22]{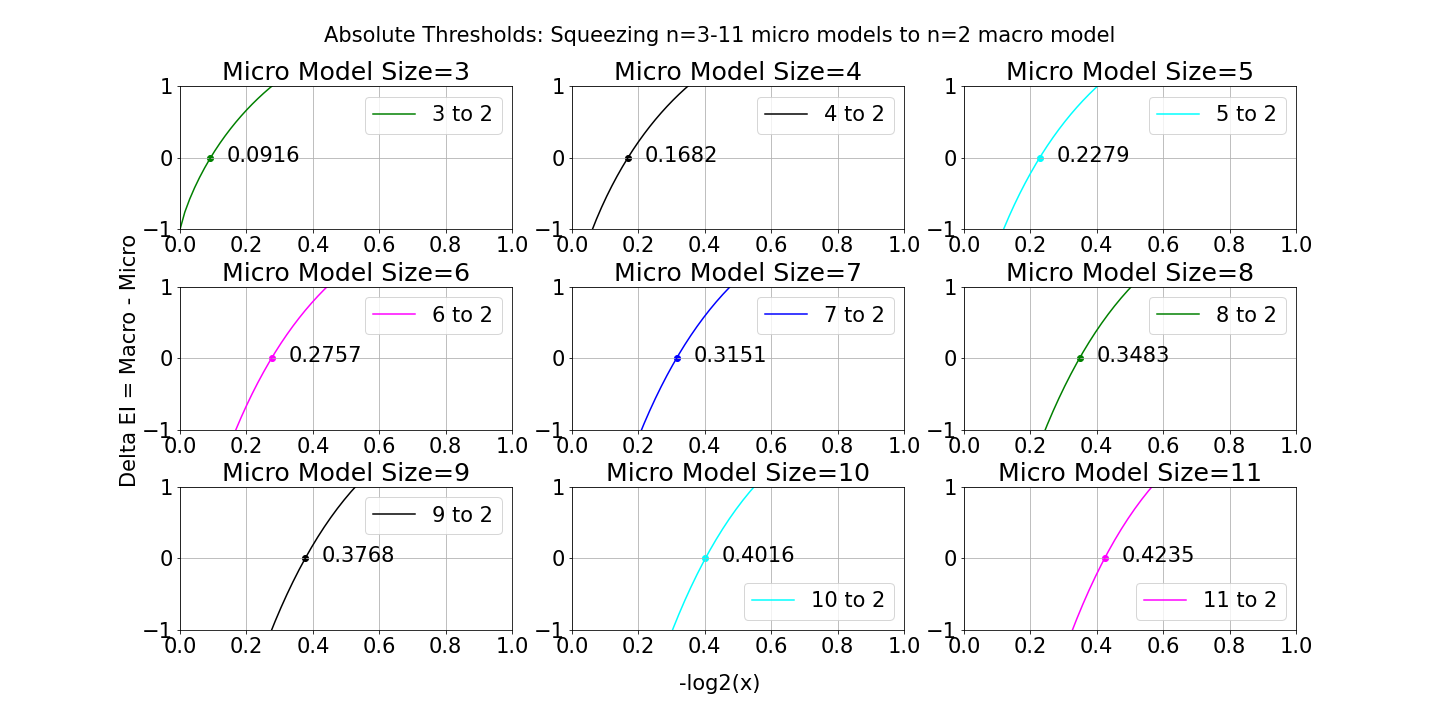} 
    \caption{Absolute Thresholds indicate uncertainty limitations of nine microscopic models with scales $n=3$, $n=4$, $n=5$, $n=6$, $n=7$, $n=8$, $n=9$, $n=10$, and $n=11$, respectively, when coarse-graining them into a macroscopic model with $n=2$ scale. The value of Absolute Threshold relates to microscopic scale $n(\text{CM\_m})$.} 
    \label{Fig_Abs_Thres_n_2} 
\end{figure}

\begin{figure}[t] 
    \centering 
    \includegraphics[scale=0.22]{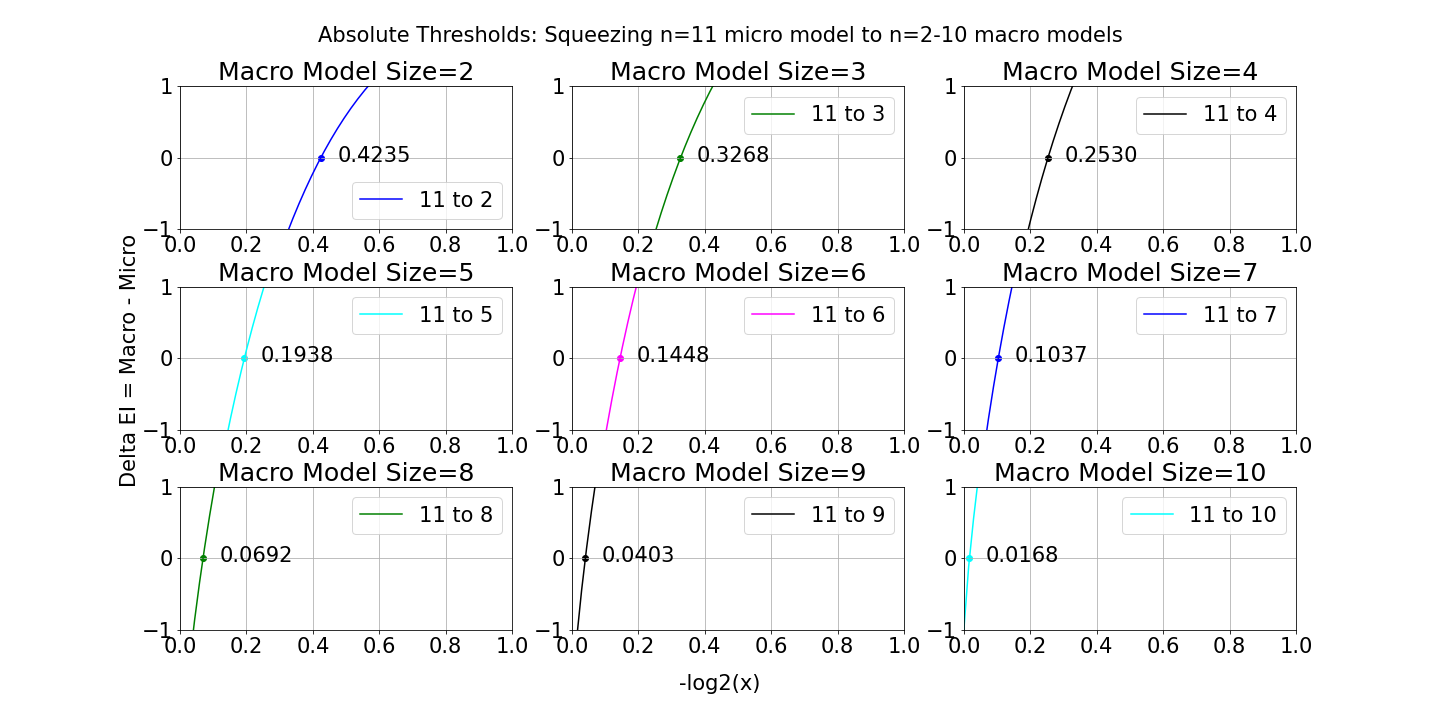} 
    \caption{Absolute Thresholds indicate uncertainty limitations of a microscopic model with $n=11$ scale when coarse-graining it into nine macroscopic models with $n=2$, $n=3$, $n=4$, $n=5$, $n=6$, $n=7$, $n=8$, $n=9$, and $n=10$, respectively. The value of Absolute Threshold relates to macroscopic scale $n(\text{CM\_M})$.} 
    \label{Fig_Abs_Thres_11_n} 
\end{figure}

The CM\_m provides a more careful description of causal relationships within the system than the CM\_M, as it has more variables to hold more system information. However, the CM\_m not only maximizes amounts of information but is also more fragile to errors and noise inherent in the system. There are instances where the microscopic $EI(\text{CM\_m})$ is lower than the macroscopic $EI(\text{CM\_m})$, and a mapping function exists within the potential space $\mathcal{F}$ that enables the CE by coarse-graining the CM\_m into a CM\_M with higher EI.
Hence, when the goal is to achieve CE through coarse-graining the CM\_ms into the CM\_ms, the primary requirement is that microscopic $EI(\text{CM\_m})$ must be decreased by model's negative characteristics, such as the uncertainty represented as $-\log_2(x)$, to ensure that a CM\_M with a higher $EI(\text{CM\_M})$ exists in the space of system's all causal models. The analysis in this section focused on the CE condition of CM's uncertainty. Consequently, we set $deg\_vector=[1,\ 1]$, meaning that every CM tested by experiments in this section is symmetric, and their maximal EI is not influenced by $degeneracy$.

Figures \ref{Fig_Abs_Thres_n_2} and \ref{Fig_Abs_Thres_11_n} show the critical CE condition on CM\_m's uncertainty, namely as Absolute Threshold (AT). The $\Delta EI$ curves in the two figures are to demonstrate that CE is achievable by satisfying the indicator, $\Delta EI = EI(\text{CM\_M}) - EI(\text{CM\_m}) > 0$, when the microscopic uncertainty surpasses the values of ATs. These results show that the CM\_ms must exhibit a certain level of uncertainty, surpassing the AT's values calculated by Equation \ref{reg_cqe} based on CM\_m and CM\_M taking part in the coarse-graining strategy, to ensure that a more effective CM\_M can be found in the space of system's CMs.

\begin{figure}[t] 
    \centering 
    \includegraphics[scale=0.32]{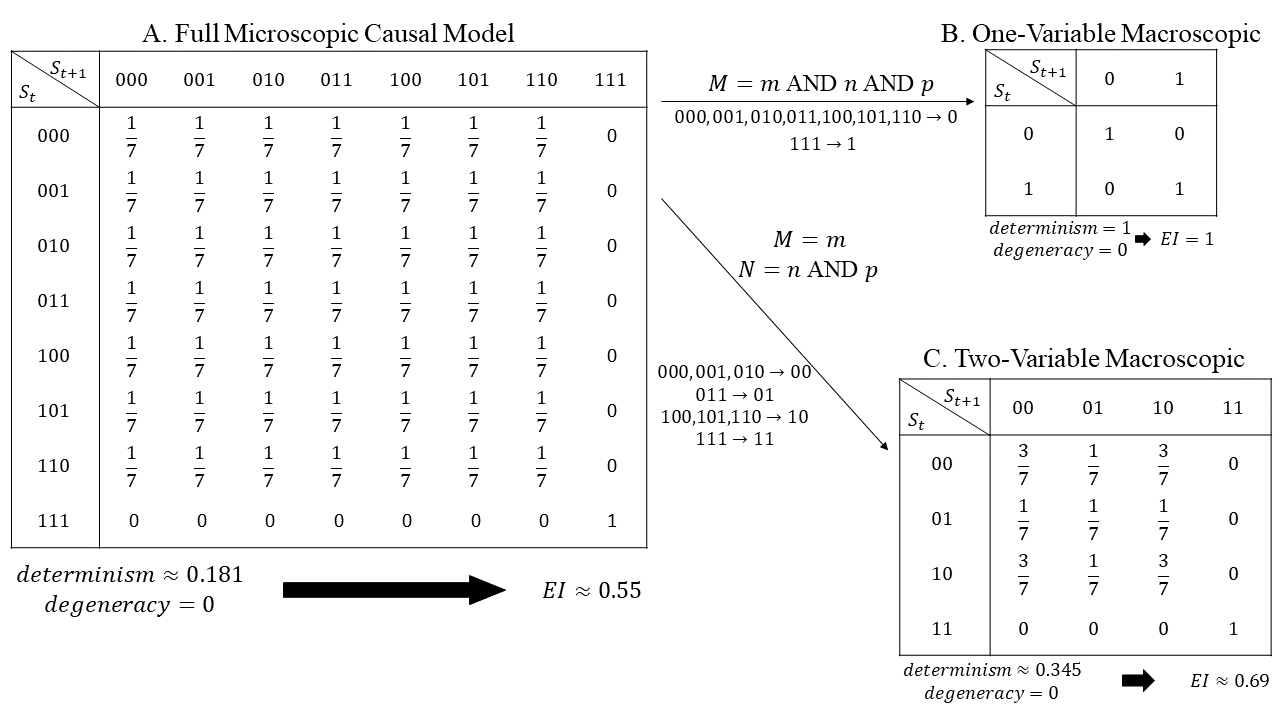} 
    \caption{Example of coarse-graining a full microscopic CM into the macroscopic, whose scale is slightly smaller than the microscopic. Finally, the performance of coarse-graining strategy, whose target is four-state model, is overwhelmed by coarse-graining the CM\_m into two-state model.} 
    \label{coarse-exp1} 
\end{figure}

However, eliminating the entire uncertainty in the CM\_m to obtain a deterministic CM\_M through coarse-graining strategies in real scenarios is challenging. As shown in Figure \ref{coarse-exp1}, coarse-graining the eight-state microscopic model into a four-state model only increases the EI from $0.55$ to $0.69$ bits. There is still residual uncertainty in the causal structure of the model with four states. Hence, the performance of this strategy is more inefficient than coarse-graining the microscopic into the two-state model, as displayed in Figure \ref{coarse-exp1}.

\begin{figure}[b] 
    \centering 
    \includegraphics[scale=0.3]{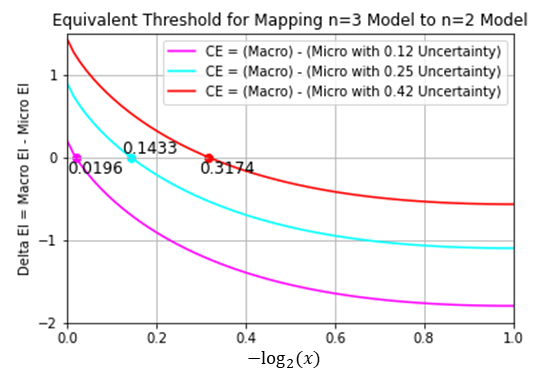} 
    \caption{Equvalent Thresholds indicate uncertainty limitations of a macroscopic model with $n=2$ scale when coarse-graining three $n=3$'s microscopic model with  $0.12\text{ bits}$, $0.25\text{ bits}$, and $0.42\text{ bits}$ uncertainties, respectively, into the macroscopic.} 
    \label{ets} 
\end{figure}

Consequently, it is imperative to consider the CE condition regarding CM\_M's uncertainty. Using the number $n(\text{CM\_M})$ of CM\_M's variables and the uncertain microscopic model's $EI(\text{CM\_m})$ as inputs of Algorithm \ref{solver}, Figure \ref{ets} shows three examples of coarse-graining three eight-state CM\_ms into a four-state CM\_M when microscopic uncertainties are $0.12$ bits, $0.25$ bits, and $0.42$ bits, respectively.
According to the quantifications by Algorithm \ref{solver}, the macroscopic uncertainties required to achieve $EI(\text{CM\_M})=EI(\text{CM\_m})$ represent lower limitations of CM\_M's uncertainties, as called Equivalent Thresholds (ETs), whose values are smaller than the uncertainties of three CM\_ms. Those results shown in Figures \ref{Fig_Abs_Thres_n_2}, \ref{Fig_Abs_Thres_11_n}, and \ref{ets} quantitatively valid the first CE condition: CE follows the reduction in the CM's uncertainty, which the magnitude exceeds the difference between AT and ET, denoted as $\Delta Uncertainty > AT - ET$, when coarse-graining a stochastic microscopic model into a more deterministic macroscopic model.

\section{Quantifying the critical conditions of CE: The necessity of considerations about asymmetry decrement}
\label{results_asymmetry}

\begin{figure}[t] 
    \centering 
    \includegraphics[scale=0.35]{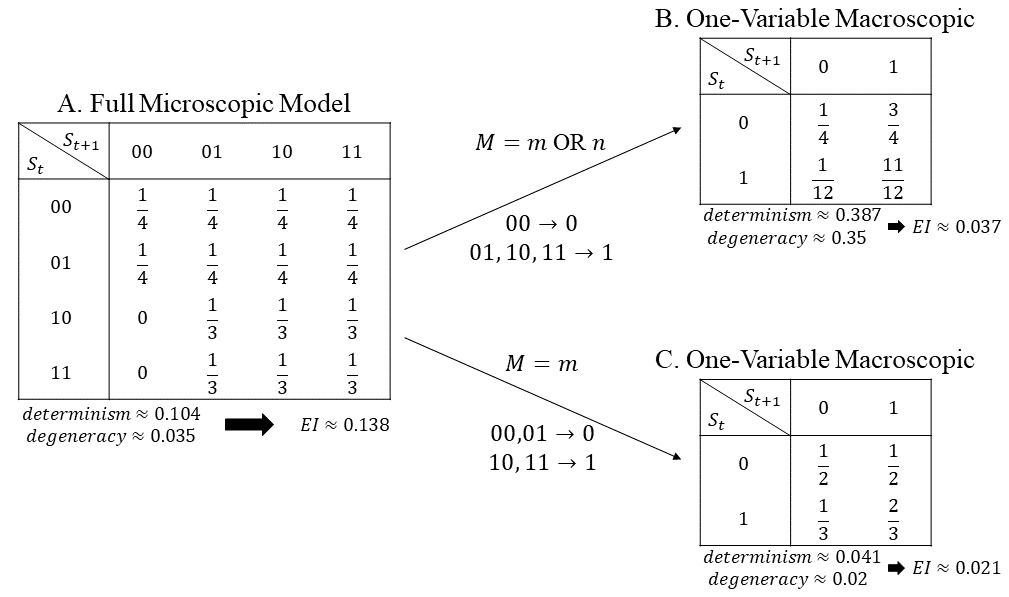} 
    \caption{Two coarse-graining strategies transform four-state microscopic model into two-state macroscopic model. The first aims to decrease model's uncertainty for the CE, and another focuses on asymmetry reduction for the CE.} 
    \label{exp_deg} 
\end{figure}

Equation \ref{EI_decop} \cite{hoel2017map} highlights that the impact of asymmetry on the model's EI is distinct from that of uncertainty. Asymmetry is an inherent characteristic of CMs themselves, and it directly lowers the EI maximization discretely. Hence, it is also crucial to address how to ameliorate or eliminate asymmetry in pursuit of CE by coarse-graining a microscopic model into a macroscopic model.

Figure \ref{exp_deg} provides an example of projecting a four-state microscopic model into two-state macroscopic models through two coarse-graining strategies. Although the first strategy successfully reduces the CM\_m's uncertainty, CE does not occur as that $degeneracy$ increases. Conversely, the second strategy effectively alleviates the CM\_m's asymmetry, but CE disappears with $determinism$ drops. This result highlights the coarse-graining strategies for CE need to optimize both uncertainty and asymmetry of CMs simultaneously. This situation complicates the coarse-graining choices. Therefore, we define two preconditions to study the numerical CE conditions of asymmetric CMs.
\begin{figure}[b] 
    \centering 
    \includegraphics[scale=0.35]{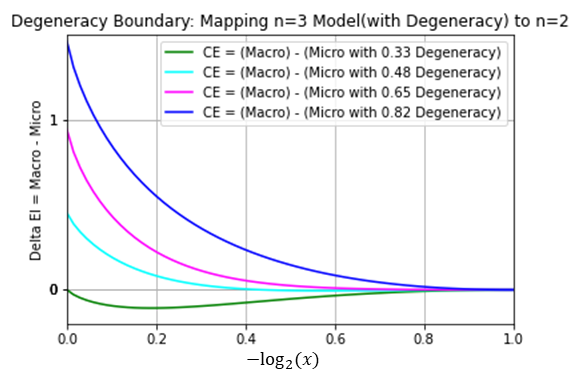} 
    \caption{Degeneracy Boundary of coarse-graining an eight-state microscopic model into a four-state macroscopic model without uncertainty and asymmetry. When microscopic $degeneracy$ exceeds DB value, $0.33\text{ bits}$, system's model space always has a four-state macroscopic model with greater $EI(\text{CM\_M})$.} 
    \label{db} 
\end{figure}

\begin{figure}[t] 
    \centering 
    \includegraphics[scale=0.21]{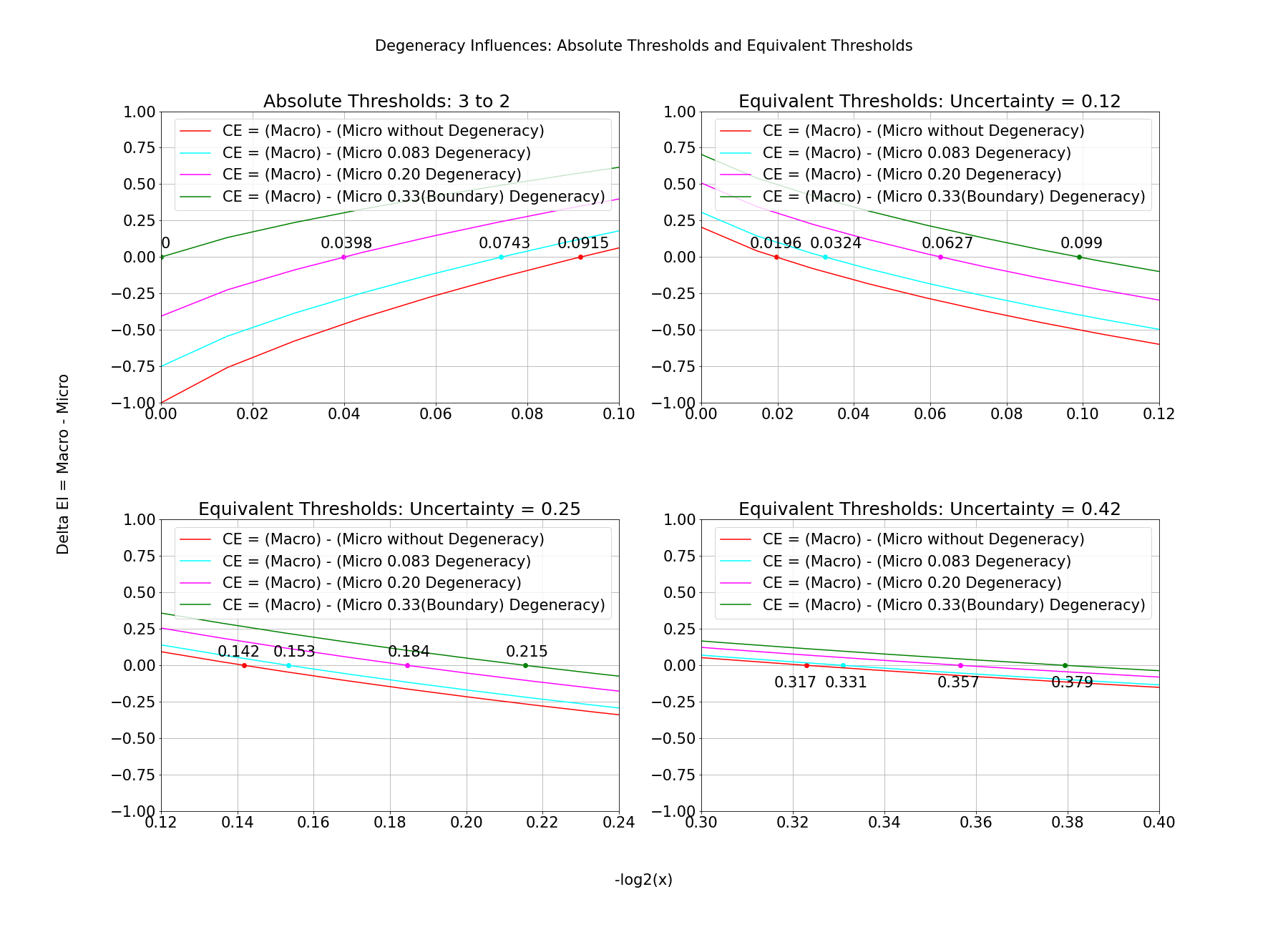} 
    \caption{Degeneracy Influence of ATs and ETs of coarse-graining an asymmetric eight-state microscopic model into a symmetric four-state macroscopic model. Following increment of $degeneracy$ of microscopic models, AT's values decrease from $0.0915\text{ bits}$ to $0\text{ bits}$ when microscopic $degeneracy$ equals the DB. ET's values increase, which indicates that the space of CE-available macroscopic models with $EI(\text{CM\_M})>EI(\text{CM\_m})$ is expanded by microscopic $degeneracy$.} 
    \label{di} 
\end{figure}
Firstly, we assume the origins of coarse-graining strategies are asymmetric but deterministic models. According to Equation \ref{EI_decop} \cite{hoel2017map}, asymmetry is an intrinsic characteristic of models to lower the EI maximization. When the CM\_m is asymmetric enough, even if it is deterministic, there exists a CM\_M within the space of all the system's CMs, whose $EI(\text{CM\_M})$ is bigger than $EI(\text{CM\_m})$. We define Degeneracy Boundary (DB) to indicate the $degeneracy$ that decreases the CM\_m's maximization $EI^{max}(\text{CM\_m})$ to equal the CM\_M's maximization $EI^{max}(\text{CM\_M})$. 

Using Equations \ref{reg_det} and \ref{reg_cqe}, we can calculate the DB for coarse-graining microscopic models into macroscopic models. In Figure \ref{db}, coarse-graining a deterministic but asymmetric CM\_m with $n=3$ into a $n=2$ CM\_M without uncertainty and asymmetry has a DB of $0.33\text{ bits}$, calculating from $determinism(\text{CM\_m}) = 1$ and $\frac{EI(\text{CM\_M})}{n(\text{CM\_m})} = \frac{2}{3}$. As a result, when the $degeneracy$ of microscopic model exceeds the DB, coarse-graining strategies can focus on ameliorating or eliminating asymmetry of microscopic models when squeezing them into macroscopic models.

Secondly, when the CM\_m is uncertain and asymmetric, coarse-graining strategies must simultaneously optimize $determinism$ and $degeneracy$. Therefore, as shown in Figure \ref{di}, when the $degeneracy$ of an eight-state CM\_m does not surpass the DB, we can apply our framework to calculate ATs and ETs to prove the uncertainty within asymmetric cases still constrains the CE occurrence. In this experiment, we still assume that CM\_M is symmetric to emphasize the uncertainty's impact on CE, and the results in Figure \ref{di} demonstrate that the ETs are still smaller than the CM\_m's uncertainties. Therefore, while alleviating or eliminating CM's asymmetry, coarse-graining strategies must focus on reducing uncertainty to achieve CE.

\section{Tried Extensions of Quantification Framework and Possible Improvements}
\label{discussion}
In Sections \ref{results_uncertainty} and \ref{results_asymmetry}, we have presented our experimental results derived from the quantification framework to verify numerical constraints that coarse-graining strategies must satisfy when the CE occurs. Firstly, for the uncertainty of CMs, the reduction caused by the coarse-graining is required to exceed the difference between the CM\_m's ATs and the CM\_M's ETs, indicated as $\Delta Uncertainty > AT - ET$. This constraint is the necessary CE condition even if the CM's EI is influenced by the asymmetry measured by $degeneracy$. Secondly, for the asymmetry of CMs, it is essential to alleviate or eliminate the $degeneracy$ of CM\_m during the coarse-graining process. Remarkably, the elimination of asymmetry will bring the CE when the CM\_m's $degeneracy$ surpasses the value of DB, as the maximum EI of the CM\_m is less than the biggest effectiveness $H(I_D(\text{CM\_M}))$ of CM\_M.

In this section, we will discuss our framework's potential applications for designing coarse-graining strategies for CE and training neural networks to predict the CE. The discussions are based on previous research \cite{hoel2013quantifying, rosas2020reconciling} or the results of our extensive experiments. Meanwhile, the analyses of possible improvements in our quantification framework are provided to implement these ideas in the future to support the CE-relative studies.

\subsection{Potential Application: The Design of Coarse-graining Strategies for CE}
\label{extension1}
In previous studies, Hoel \cite{hoel2013quantifying, hoel2017map} defined coarse-graining strategies as the aggregation or ignorance of CM\_m variables to obtain corresponding macroscopic variables, resulting in a more efficient CM\_M. Our work interprets the "aggregation" operation as a mathematical relationship between microscopic and macroscopic variables established by logical operations. As shown by several experimental examples, we have employed expressions, such as $M= m\text{ AND }n\text{ AND }p$ in Figure \ref{coarse-exp1} or $M = m\text{ OR }n$ in Figure \ref{exp_deg}, to represent our applied coarse-graining strategies. Moreover, as our quantification framework introduced controlled uncertainty into synthetic TPMs through the Variables' Mechanism, i.e., VAM, we believe this framework can potentially guide the design of coarse-graining strategies for CE. Here, we discuss this potential application scenario for our quantification framework based on Causal Decoupling, as proposed by Rosas et al. \cite{rosas2020reconciling}. We propose potential improvements by analyzing the preliminary experimental results to realize our idea in future works.

\begin{figure}[b] 
    \centering 
    \includegraphics[scale=0.30]{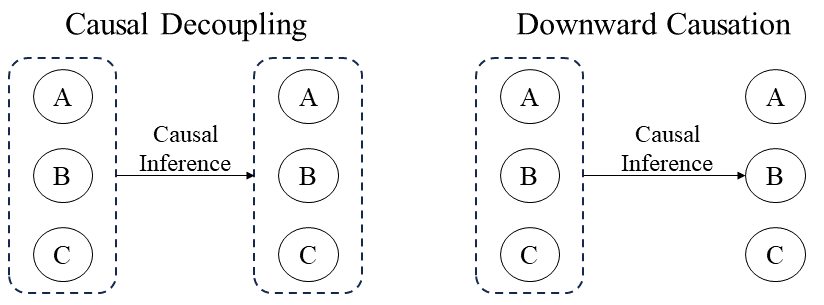} 
    \caption{Causal Decoupling and Downward Causation are two subtypes of CE defined by Rosas et al. \cite{rosas2020reconciling}.} 
    \label{CE_sub} 
\end{figure}

To avoid the limitations and irrationalities that Hoel's MED hypothesis introduced in CE theory, Rosas et al. \cite{rosas2020reconciling} adopted the PID \cite{williams2010nonnegative,james2018unique} theory to define CE in their research. They quantified indicators of CE through the variation in Partial Information. Although this approach was excluded from our quantification framework due to its high computational complexity \cite{williams2010nonnegative,james2018unique}, it offers valuable definitions. Rosas et al. \cite{rosas2020reconciling} distinguished two CE subtypes, Causal Decoupling and Downward Causation, as illustrated in Figure \ref{CE_sub}. Causal Decoupling signifies a causal relationship between the overall state of CM's variables and their collective future state. For example, regardless of which one of the microscopic variables, A, B, and C, is currently activated, one or more of them will be activated in the future. Those microscopic variables are more suitable for merging into a macroscopic variable, such as $\alpha = A\text{ AND }B\text{ AND }C$, to enhance the clarity and effectiveness of causal relationships in the CM. Downward Causation denotes a causal relationship between the overall state of CM's variables and the future state of one specific variable within them. For instance, regardless of which one of the microscopic variables, A, B, and C, is currently activated, variable B will deterministically be activated in the next time step. This phenomenon also supports an alternative coarse-graining approach to aggregate microscopic variables. The influence of variable B in determining the value of the macroscopic variable is as significant as possible in such an approach.

\begin{figure}[t] 
    \centering 
    \includegraphics[scale=0.25]{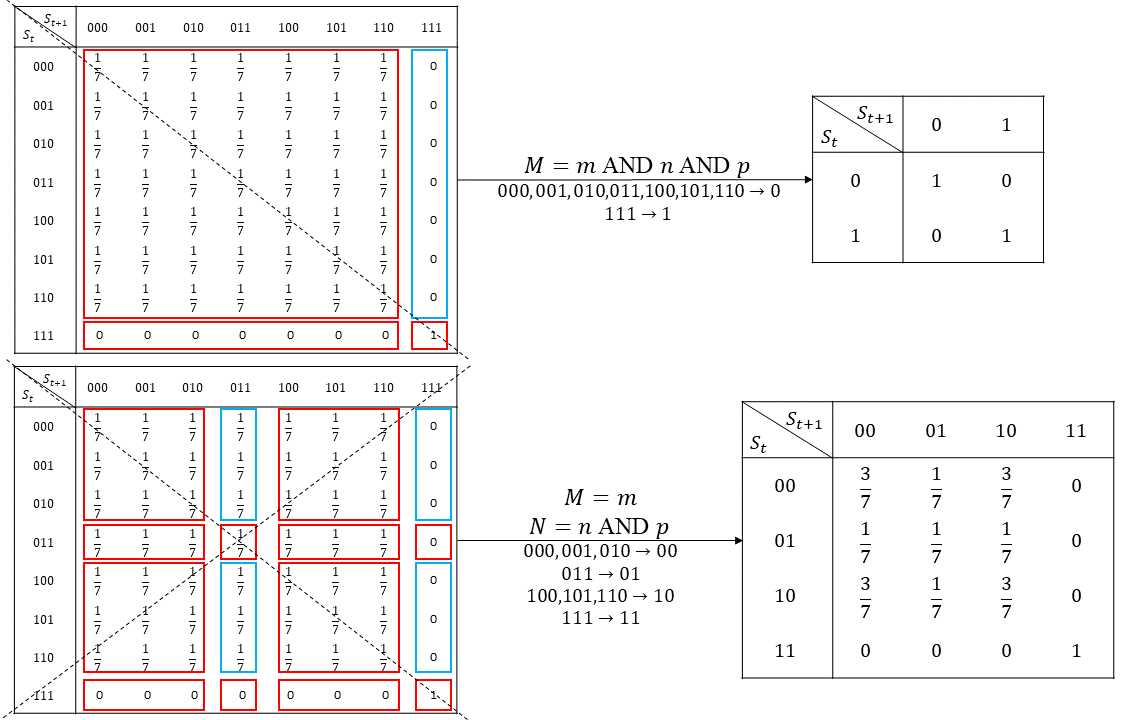} 
    \caption{Circling subsets for coarse-graining an eight-state microscopic model into two-state and four-state macroscopic models, respectively.} 
    \label{subsets} 
\end{figure}

\begin{figure}[b] 
    \centering 
    \includegraphics[scale=0.25]{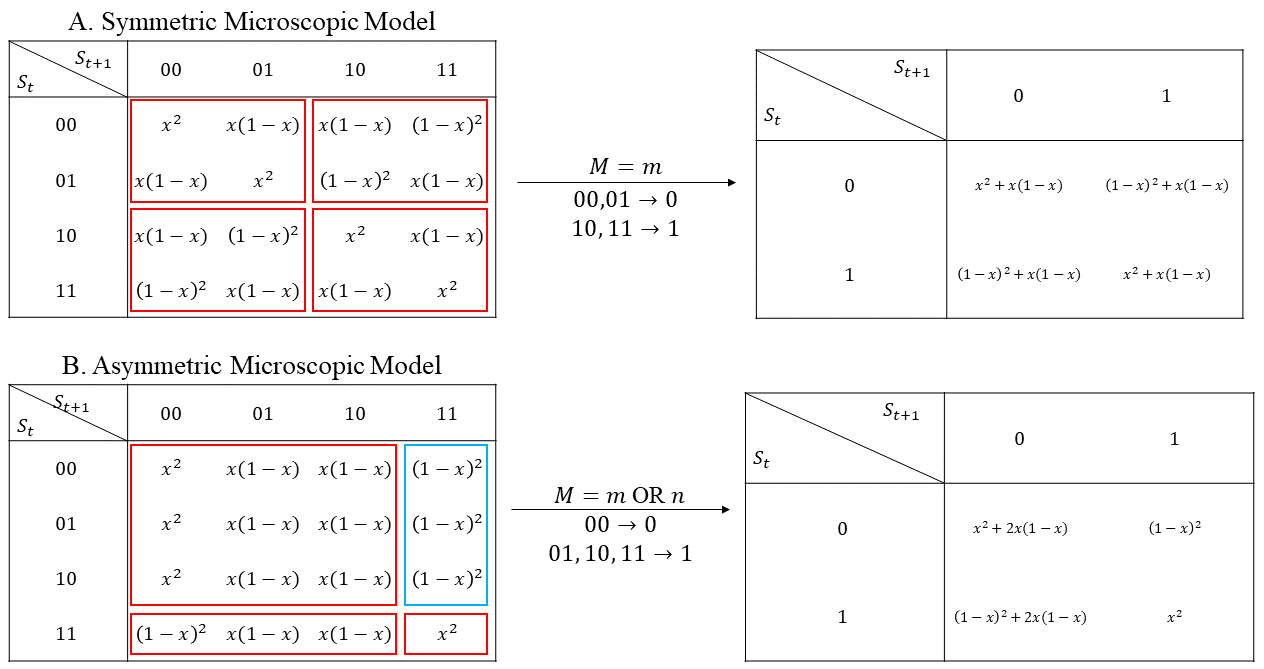} 
    \caption{Design coarse-graining strategies for a symmetric and stochastic two-state CM and an asymmetric and stochastic two-state CM based on framework's synthetic TPMs, whose elements are polynomials of $x$.} 
    \label{design} 
\end{figure}

While we can intuitively represent coarse-graining strategies using mathematical relationships defined by logical operations, the corresponding implementation of coarse-graining to the TPM of CM still requires establishing a mapping relationship between the states of CM\_m and CM\_M. If we visualize the coarse-graining's state mapping approaches on the CM\_m's TPM by subsets, as shown in Figure \ref{subsets}, we can observe that the state transitions in subsets along the diagonal can be described by Causal Decoupling. For instance, at the top of Figure \ref{subsets}, the coarse-graining process groups all $\frac{1}{7}$ elements around the diagonal into a red subset, representing that those state transitions exhibit Causal Decoupling. The microscopic states in this red subset, including $000$, $001$, $010$, $011$, $100$, $101$, and $110$, have a comprehensive causal effect on the collection of corresponding future states. Consequently, these states are mapped to a macroscopic $0$ in CM\_M. For the coarse-graining strategy displayed at the bottom of Figure \ref{subsets}, the process needs to be represented by two logical expressions. Therefore, Causal Decoupling subsets on the CM\_m's TPM are on two diagonals. Around these red subsets, some additional subsets are distributed. The additional subset below indicates state transitions that comply with Causal Decoupling. The subset on the right can represent either Downward Causation (as shown by blue subsets in Figure \ref{subsets}) or Causal Decoupling, as illustrated by the coarse-graining at the bottom of Figure \ref{design}.

Therefore, we suppose that an effective coarse-graining strategy for CE is to build one or multiple Causal Decoupling subsets upon CM\_m's TPM, and each subset ideally contains the most amount of uncertain and asymmetric state transitions to reduce CM\_m's uncertainty and asymmetry. Since all its state transition probabilities are replaced by x polynomials, our synthetic TPMs can derive coarse-graining strategies capable of bringing CE. Following $x\in [0.5,\ 1]$, the $x^n$ in the TPM indicates the maximum probability of a specific current state transitioning to a future state. Consequently, more inclusion of $x^n$ in Causal Decoupling subsets, as illustrated in Figure \ref{design}, allows aggregations of microscopic states to have a more significant impact on increasing EI.

Although we have applied preliminary experiments to validate this idea, the results yielded unfavorable conclusions. Only a few coarse-graining strategies designed on synthetic TPMs could make CE occur when applying them to real TPMs with the same number $n$ of CM's parameter. Furthermore, even for these valid coarse-graining methods, their effectiveness can not be directly validated through synthetic TPMs. We attribute this situation to the dissimilarity between synthetic TPMs and real TPMs. The comparison of TPMs in Figures \ref{subsets} and \ref{design} reveals that the uncertainty in real TPM is discretely distributed across state transition probabilities, while our synthetic TPM, due to the design of a single parameter $x$, results in uniformly distributed uncertainty across overall state transitions. Therefore, to enhance the realism of our quantification framework's controllable TPM, we propose future exploration using multiple parameters $X={x_1,\ x_2, \cdots,\ x_i}$, allowing each group of current and future states to represent a specific level of uncertainty. Through making our synthetic TPM closer to a real TPM, calculation the growth of EI contributed by each Causal Decoupling subset leads us to design coarse-graining strategies for CE purposefully.

\subsection{Potential Application: Neural Networks for Predicting CE Conditions}
\label{extension2}

Based on our newly established quantification algorithm in Equations \ref{reg_det} and \ref{reg_cqe} (demonstrated by Algorithm \ref{solver} in Appendix \ref{extensions}), this process also depends on the complex mechanism for generating synthetic TPMs. The efficiency of our quantification framework relies on the number of iterations required to identify the precise CE conditions. The more iterations needed, the longer the time for calculating numerical constraints for coarse-graining strategies. Since the principle of our quantification algorithm is selecting a pair of samples from the possible value spaces of $x$ and $deg\_vector$ in each iteration to discover a specific group $[x, deg\_vector]$ as the critical conditions for CE, the number of iterations performed by the framework is contingent upon the number of potential samples. To derive an accurate value of $x$ as the uncertainty condition required for CE, we sampled 1000 values from the $x$ range,  $[0.5, 1]$, as possible samples of CM's uncertainty. For all possible samples of CM's asymmetry represented by binary arrays as $deg\_vector$s, Table \ref{complexity_with_n} provides the relationship between the number of possible arrays and the number $n$ of CM variables. For example, to identify the critical CE conditions necessary for coarse-graining a CM\_m with seven variables to a CM\_M with $EI_{target}$, there are 4097 $deg\_vector$s for testing by our quantification framework. Therefore, the framework requires to find that specific pair from $4\times 10^6$ possible $[x, deg\_vector]$ samples. This process requires substantial computational resources and time.

\begin{table}[h]
\centering
\caption{The Complexity of Quantification Algorithm in Figure \ref{quan_process2}.}
\begin{tabular}{|c|c|}
\hline
Number of Variables $n$ & Complexity (Number of $deg\_vector$s) \\ \hline
1                   & 0                     \\ \hline
2                   & 5                     \\ \hline
3                   & 17                    \\ \hline
4                   & 65                    \\ \hline
5                   & 257                   \\ \hline
6                   & 1025                  \\ \hline
7                   & 4097                  \\ \hline
\end{tabular}
\label{complexity_with_n}
\end{table}

To address the computational complexity issue of our quantification framework, we attempt to leverage data generated by the current quantification mechanism to construct a dataset for training Artificial Neural Networks (ANN). Those data include the $n$, $determinism$, $degeneracy$, and $EI_{test}$ of tested CM, and the testing sample $[x, \ deg\_vector]$. We utilize this dataset to train a regression ANN. Beneficial to the powerful predictive capabilities of neural networks \cite{bishop2006pattern, goodfellow2016deep}, the trained ANN can directly predict the critical condition for CE from the input values of $EI_{target}$ and $n_{test}$. Appendix \ref{regression} shows the introductions of the dataset, the ANN, and relevant settings for training the network.

\begin{figure}[t] 
    \centering 
    \includegraphics[scale=0.65]{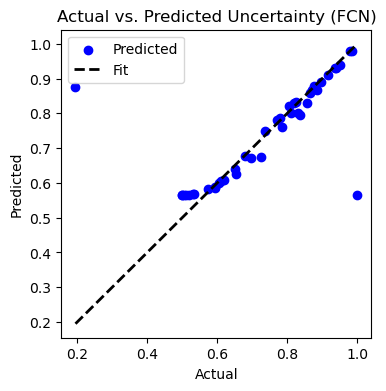} 
    \caption{The performance of trained regression ANN on the testing set. The details of the dataset, the ANN, and training settings are described in Appendix \ref{regression}.} 
    \label{exp_conversion} 
\end{figure}

We employ experimental results of critical CE conditions and the corresponding inputs, $EI_{target}$s and $n_{test}$s, to be a part of the test set to validate the trained ANN. Additionally, another subset of the test set includes data, denoted as $EI_{real}$, $n_{real}$, $x_{real}$, and $deg\_vector_{real}$, derived from TPMs in Hoel's papers \cite{hoel2013quantifying,hoel2017map}, where the values of $x_{real}$ and $deg\_vector_{real}$ are obtained by using $EI_{real}$ and $n_{real}$ as inputs for Algorithm \ref{solver}. These testing data can assess the network's generalization capability to samples beyond the training distribution to verify the practicality of ANN. Figure \ref{exp_conversion} shows the predictions of the trained ANN on our test set. In Figure \ref{exp_conversion}, the dashed line represents the ground truth of predictions, and the blue dots indicate the network's predictions on the test set. Most blue dots cluster around the dashed line, indicating the effectiveness of the ANN predictions on test samples. However, the network's predictions could be more satisfactory on some samples derived from Hoel's TPMs\cite{hoel2013quantifying,hoel2017map}, as evidenced by the blue dots that deviate from the dashed line. We realize that the main reason for this deficiency is still the low similarity between the synthetic TPMs and the real TPMs, which results in the distribution of data generated by our framework being different from the distribution of values of the real TPM's characteristics in practical scenarios.

Instead of the cause of the single parameter $x$ in subsection \ref{extension1}, the parameter $deg\_vector$ is another contributor concerning the dissimilarity. Firstly, the $[x,\ deg\_vector]$ pairs obtained through Algorithm \ref{solver} could be inaccurate. This notion arises from the observation of differences between synthetic $determinism$ and $degeneracy$ and their real counterparts, $determinism_{real}$ and $degeneracy_{real}$, during the conversion of Hoel's TPMs \cite{hoel2013quantifying,hoel2017map} to synthetic TPMs. We directly utilize TPM's $EI_{real}$ as a condition to constrain the similarity between synthetic TPMs and Hoel's TPMs \cite{hoel2013quantifying,hoel2017map}. However, in experiments, we discovered significant but equivalent errors between synthetic $determinism$ and $degeneracy$ and $determinism_{real}$ and $degeneracy_{real}$, which were not reflected in the gap between $EI_{real}$ and $EI_{test}$ because of Equations \ref{EI_decop} and \ref{effectiveness}. This situation implies that the test samples derived from Hoel's TPMs \cite{hoel2013quantifying,hoel2017map} could be imprecise. Secondly, within our quantification framework, the TPM's asymmetry is implemented fixedly based on $deg\_vector$. Similar to the challenges arising from the approach for implementing uncertainty through the single variable $x$, this straightforward method introduces dissimilarities to cause a gap between the distribution of the data generated by our quantification framework and the distribution of the relevant data from real TPMs.

We expect to employ a multi-objective optimization approach \cite{thu2008multi} to address the first inaccurate shortage. It allows the errors of both $determinism$ and $degeneracy$ to be constraints for validating the similarity between synthetic TPMs and real TPMs. Therefore, this methodology can mitigate significant differences between the two inherent coefficients of the TPM and enhance our ANN using multi-task learning \cite{liu2016deep}, whose foundation is multi-objective optimization \cite{thu2008multi}. The current ANN specializes in regression tasks, predicting the continuous variable $x$ as the CE condition on uncertainty. The $deg\_vector$ representing the CE condition on asymmetry is obtained through an auxiliary Algorithm \ref{deg_gen} based on the predictions of the ANN. With multi-task learning \cite{liu2016deep}, the ANN can simultaneously learn regression tasks for predicting continuous variable $x$ and classification tasks for discrete variable $deg\_vector$, enabling the direct prediction of numerical CE conditions from input values $EI_{target}$ and $n_{test}$. For the second dissimilarity deficiency, we consider to design new asymmetry representation in the future to enhance the similarity between synthetic TPMs and real TPMs. Despite the existing challenges in addressing the computational complexity of our current quantification framework, this extensive experiment is a preliminary demonstration of potential application scenarios for our work in future research, which could serve as a fundamental training data for CE in complex DL networks \cite{goodfellow2016deep}.

\section{Conclusions}
\label{conclusion}
In conclusion, our study has presented a quantification framework to numerically demonstrate that CE depends on reducing a causal model's (CM's) uncertainty and asymmetry by coarse-graining strategies. We have established a CQE equation to quantify precise CE conditions on uncertainty and asymmetry of CMs, i.e., the microscopic causal model (CM\_m) and the macroscopic causal model (CM\_M) taking part in the coarse-graining process. Our experimental results in Sections \ref{results_uncertainty} and \ref{results_asymmetry} have shown that CM\_ms, as the source of coarse-graining strategies, must exhibit stochastic or asymmetric characteristics to meet the microscopic lower limitations represented by Absolute Thresholds (ATs) and Degeneracy Boundaries (DBs) and allow the space of all system's CMs to include a CM\_M with the higher $EI(\text{CM\_M})$ than the $EI(\text{CM\_m})$. Moreover, the upper limitations of CM\_M's uncertainties can be quantified by Equivalent Thresholds (ETs), obtained by calculating Equation \ref{reg_cqe} from the $EI(\text{CM\_m})$ and the CM\_M's variable amount $n(\text{CM\_M})$. As shown in Figures \ref{ets} and \ref{di}, the values of ETs are always smaller than the microscopic uncertainties whether the CM\_m has asymmetry or not. As a result, we have quantitatively proposed the critical CE condition: for the CE occurrence, coarse-graining strategies are required to reduce the microscopic uncertainties to satisfy $\Delta Uncertainty > AT - ET$ while lessening or eliminating the asymmetry.

Additionally, our work provided extensive experiments in \ref{discussion} to explore our quantification framework's potential application scenarios, including the design of proper coarse-graining strategies for CE occurrence and the generator of training data for applying Deep Learning (DL) networks \cite{goodfellow2016deep} to investigate CE. However, the inaccuracy and dissimilarity issues challenge our framework's practical applications. To address the inaccurate challenge in future research, we expect to apply multi-object optimization \cite{thu2008multi} and multi-task learning \cite{liu2016deep} approaches to optimize the accuracy and effectiveness of training and testing data for the Artificial Neural Network (ANN). On the other hand, it is essential to define new representations of CM's uncertainty and asymmetry for substituting $x$ and $deg\_vector$ to address the dissimilarity challenge between the synthetic and real TPMs. The work proposed by Zhang and Liu \cite{zhang2022neural} gives valuable inspiration to investigate utilizing the powerful learning abilities of DL networks \cite{goodfellow2016deep} to capture more effective representations of CM's characteristics in future studies.`

In the future, rather than focusing solely on improving the practicality of our framework, we look forward to applying concepts from causality theories to enhance DL techniques. Many researchers studying causal relationships have argued that causality invariance is conducive to developing generalization ability for DL networks \cite{yao2021survey, scholkopf2021toward}. Consequently, our future research will focus on the robust learning implemented by invariant causal relationships and CE theories with numerical constraints provided by our optimized quantification framework.

\section*{CRediT authorship contribution statement}

\textbf{Liye Jia}: Investigation, Methodology, Validation, Visualization, Writing - original draft, review \& editing. \textbf{Yutao Yue}: Research Ideas, Conceptualization, Investigation, Writing - original draft, review \& editing, Project administrator. \textbf{Fengyufan Yang}: Methodology, Validation, Visualization, Writing - review \& editing. \textbf{Ka Lok Man}: Supervision, Writing - review \& editing. \textbf{Sheng-Uei Guan}: Supervision. \textbf{Erick Purwanto}: Supervision. \textbf{Jeremy Smith}: Supervision. 

\section*{Declaration of Competing Interest}
The authors declare that they have no known competing financial interests or personal relationships that could have appeared to influence the work reported in this paper.

\section*{Acknowledgements}
This work was inspired by discussions from Swarma Club. The authors would like to thank Prof. Jiang Zhang and his team, Prof. Erik Hoel and Dr. Pavel Chvykov for their valuable help and discussions. 

This work received financial support from Jiangsu Industrial Technology Research Institute (JITRI) and Wuxi National Hi-Tech District (WND).

\section*{Appendix}
\appendix
\renewcommand{\thesection}{\Alph{section}}
\renewcommand{\thesubsection}{\thesection.\arabic{subsection}}
\renewcommand{\thesection}{\Alph{section}}
\renewcommand{\appendixname}{Appendix}
\renewcommand{\appendixtocname}{List of Appendices}
\renewcommand{\appendixpagename}{List of Appendices}

\section{Implementations of the Generation Process of Synthetic TPMs}
\label{algorithms_frame}
In this section, we provided three pseudo-codes to elucidate the detailed implementation of our quantification framework in generating synthetic TPMs with controlled uncertainty and asymmetry based on $x$ and $deg\_vector$. 

Firstly, Algorithm \ref{implemt_alg_deg} outlines the implementation for generating TPMs with controllable asymmetry, leveraging the tested CM's variable number, $n_{test}$, and the specified $deg\_vector$. This algorithm's role in our framework's TPM Generator mechanism is illustrated in Figure \ref{overview1}. Additionally, Figure \ref{deg_implemt} displays a visual example of the output of Algorithm \ref{implemt_alg_deg} when the $deg\_vector = [1,\ 3]$. Furthermore, when the specific $deg\_vector = [FD,\ \sum CD]$ meets the particular constraint, $FD > 1$ and $\sum CD > (2FD+2)$, it indicates the existence of multiple $CD$ arrays representing the current CM's various asymmetric situations, satisfying the $deg\_vector = [FD,\ \sum CD]$. In such cases, to determine precise critical CE conditions by systematically exploring all potential asymmetries in the CM, we establish an auxiliary functional block, GAP, as delineated by Algorithm \ref{gap block} in Appendix \ref{gap}.

\begin{algorithm}
\caption{Implementation of Asymmetry Defined by $deg\_vector$}
\label{implemt_alg_deg}
\begin{algorithmic}
\State // \textit{inputs and output}
\State \textbf{Input:} $n_{test}$ and $deg\_vector=[FD,\ \sum CD]$
\State \textbf{Output:} TPM\_set \text{ \# A set contains all possible asymmetric TPMs}
\State // \textit{implement}
\State initial\_TPM = identity($2^n$)  \text{ \# Initialize a symmetric TPM with $2^n$ of states}
\State \textbf{if} $FD$ == 1 and $\sum CD$ == 1: \text{ \# When the setting is no asymmetry}
\State\quad TPM\_set.append(initial\_TPM)  \text{ \# Save and output the symmetric TPM}  
\State \textbf{if} $FD$ == 1 and $\sum CD$ > 1: \text{ \# When a single degenerate state $s^f_{t+1}$}
\State\quad copied\_row = initial\_TPM[0]
\State\quad deg\_TPM = initial\_TPM
\State\quad deg\_TPM[:\ $\sum CD$] = copied\_row  \text{ \# Implement the asymmetric TPM}
\State\quad TPM\_set.append(deg\_TPM)  \text{ \# Save and output the asymmetric TPM}  
\State \textbf{if} $FD$ > 1 and $\sum CD$ < $(2*FD+2)$: \text{ \# When a single redundant array $CD$}
\State\quad CD = zeros($FD$)  \text{ \# Initialize the array of redundant current states}
\State\quad CD[0] = ceil($\frac{\sum CD}{FD}$) \text{ \# Let the first degenerate state accept the most redundancy}
\State\quad CD[1:] = floor($\frac{\sum CD}{FD}$) \text{ \# Let other degenerate states accept the least redundancy}
\State\quad start = 0 and end = CD[0]  \text{ \# Implement the asymmetric TPM}
\State\quad \textbf{for} num in CD:
\State\quad\quad copied\_row = initial\_TPM[start]
\State\quad\quad deg\_TPM = initial\_TPM
\State\quad\quad deg\_TPM[start:end] = copied\_row
\State\quad\quad start = end
\State\quad\quad end += num
\State\quad TPM\_set.append(deg\_TPM)  \text{ \# Save and output the asymmetric TPM}  
\State \textbf{if} $FD$ > 1 and $\sum CD$ > $(2*FD+2)$: \text{ \# When multiple redundant arrays}
\State\quad CD = zeros($FD$)  \text{ \# Initialize the array of redundant current states}
\State\quad CD[0] = $\sum CD$ - 2*$FD$ \text{ \# Let the first degenerate state accept the most redundancy}
\State\quad CD[1:] = 2 \text{ \# Let other degenerate states accept the least redundancy}
\State\quad $\Delta$ = GAP(FD, CD[0] - ceil($\frac{\sum CD}{FD}$)) \text{ \# Generate all possible changes for CD}
\State\quad \textbf{for} $\delta$ in $\Delta$:
\State\quad\quad possible\_CD = CD + $\delta$
\State\quad\quad \textbf{if} $\sum \delta$ == 0:
\State\quad\quad\quad \textbf{for} num in possible\_CD:
\State\quad\quad\quad\quad copied\_row = initial\_TPM[start]
\State\quad\quad\quad\quad deg\_TPM = initial\_TPM
\State\quad\quad\quad\quad deg\_TPM[start:end] = copied\_row
\State\quad\quad\quad\quad start = end
\State\quad\quad\quad\quad end += num
\State\quad\quad\quad TPM\_set.append(deg\_TPM)  \text{ \# Save and output the asymmetric TPM}
\end{algorithmic}
\end{algorithm}

Secondly, Algorithm \ref{trans_st2va} describes the second generation step of synthetic TPMs within our quantification framework. As shown in Figure \ref{overview1}, this algorithm accepts the deterministic TPM provided by Algorithm \ref{implemt_alg_deg} and two framework parameters, $n_{test}$ and $x$, as inputs to generate a stochastic VAM based on Hypotheses \ref{hypo_state2variable} and \ref{hypo_variable}, introducing controllable uncertainty denoted as $-\log_2(x)$ into synthetic TPMs. The mechanism ensures that the synthetic TPMs adhere to the definitions. The sequence of transformations from A to B and then to C, illustrated in Figure \ref{x_implemt}, serves as an example to illustrate the operational process of Algorithm \ref{trans_st2va}.
\ \\

\begin{algorithm}
\caption{States2Vars Transformation}\label{trans_st2va}
\begin{algorithmic}
\State // \textit{input and initialize}
\State \textbf{Input:} $n_{test}$ and TPM and $x$
\State \textbf{Output:} VAM \text{ \# Stochastic VAM}
\State // \textit{implement}
\State init\_VAM = zeros(($n$, $2^n$)) \text{ \# Initialize the VAM}
\State \textbf{for} row in TPM: \text{ \# Find conditional probabilities of $v_i=1$ given every current state}
\State\quad index\_col = row.index(1) 
\State\quad state\_f = format(index\_col, f'0\{$n$\}b') \text{ \# Future state to which current state transfers}
\State\quad probs = [int(bit) for bit in state\_f] \text{ \# Array contains conditional probabilities}
\State\quad \textbf{for} var in init\_VAM:  \text{ \# Add probabilities represented by $x$ and $1-x$ into VAM}
\State\quad\quad index\_var = init\_VAM.index(var)
\State\quad\quad \textbf{if} probs[index\_var] == 0:
\State\quad\quad\quad var[index\_col] = $1-x$
\State\quad\quad \textbf{if} probs[index\_var] == 1:
\State\quad\quad\quad var[index\_col] = $x$
\State VAM = init\_VAM  \text{ \# Save and output stochastic VAM}
\end{algorithmic}
\end{algorithm}

Thirdly, as the final step of the TPM Generation mechanism illustrated in Figure \ref{overview1}, Algorithm \ref{trans_va2st} realizes the transformation process from VAM to TPM as defined in Hypothesis \ref{hypo_state2variable}. This process yields well-defined synthetic TPMs, whose uncertainty and asymmetry are determined by the framework's parameters, $x$ and $deg_vector$. However, since both parameters have already been introduced into our quantification framework through Algorithms \ref{implemt_alg_deg} and \ref{trans_st2va}, Algorithm \ref{trans_va2st} only needs to accept the variable amount $n_{test}$ and the stochastic VAM generated by Algorithm \ref{trans_st2va} as inputs to ensure that the complete synthetic TPM aligns with the size specified by the framework parameter $n_{test}$. The transformation from C to D in Figure \ref{x_implemt} shows the operation process of Algorithm \ref{trans_va2st}.

\begin{algorithm}
\caption{Vars2States Transformation}\label{trans_va2st}
\begin{algorithmic}
\State // \textit{input and initialize}
\State \textbf{Input:} $n_{test}$ and VAM
\State \textbf{Output:} TPM \text{ \# Stochastic TPM}
\State // \textit{implement}
\State init\_TPM = zeros(($2^n$, $2^n$)) \text{ \# Initialize the TPM}
\State f\_states = [format(inte, f'0\{$n$\}b') for inte in range($2^n$)] \text{ \# Model's future states}
\State VAM\_T = transpose(VAM) \text{ \# Transpose the VAM}
\State \textbf{for} row in VAM\_T: \text{ \# Find conditional probabilities of $v_i=1$ given current state}
\State\quad index\_row = VAM\_T.index(row) 
\State\quad \textbf{for} index in range($2^n$):  \text{ \# Calculate transition probabilities from $s^c_t$ to $s^f_{t+1}$}
\State\quad\quad $\delta$ = 1 - [int(bit) for bit in f\_states[index]]
\State\quad\quad init\_TPM[index\_row][index] = prod(abs($\delta$ - row))
\State TPM = init\_TPM  \text{ \# Save and output stochastic TPM}
\end{algorithmic}
\end{algorithm}

Finally, by executing Algorithms 1, 2, and 3 in sequence, the TPM Generation mechanism in Figure \ref{overview1} is established to generate synthetic TPMs for quantifying the critical CE conditions.

\section{Derivations of Mathematical Relationship between $x$ and $determinism$}
\label{derivation_eqs}
\renewcommand\thefigure{\thesection.\arabic{figure}}
\renewcommand\thetable{\thesection.\arabic{table}}
\renewcommand{\theequation}{B\arabic{equation}}
\setcounter{figure}{0} 
\setcounter{table}{0} 
\setcounter{equation}{0}

In this section, we will explain the details of the regularities within the row distributions $row_i$ of synthetic TPMs and provide the derivation of our quantification equation, the CQE.

\begin{figure}[t] 
\centering 
\includegraphics[scale=0.3]{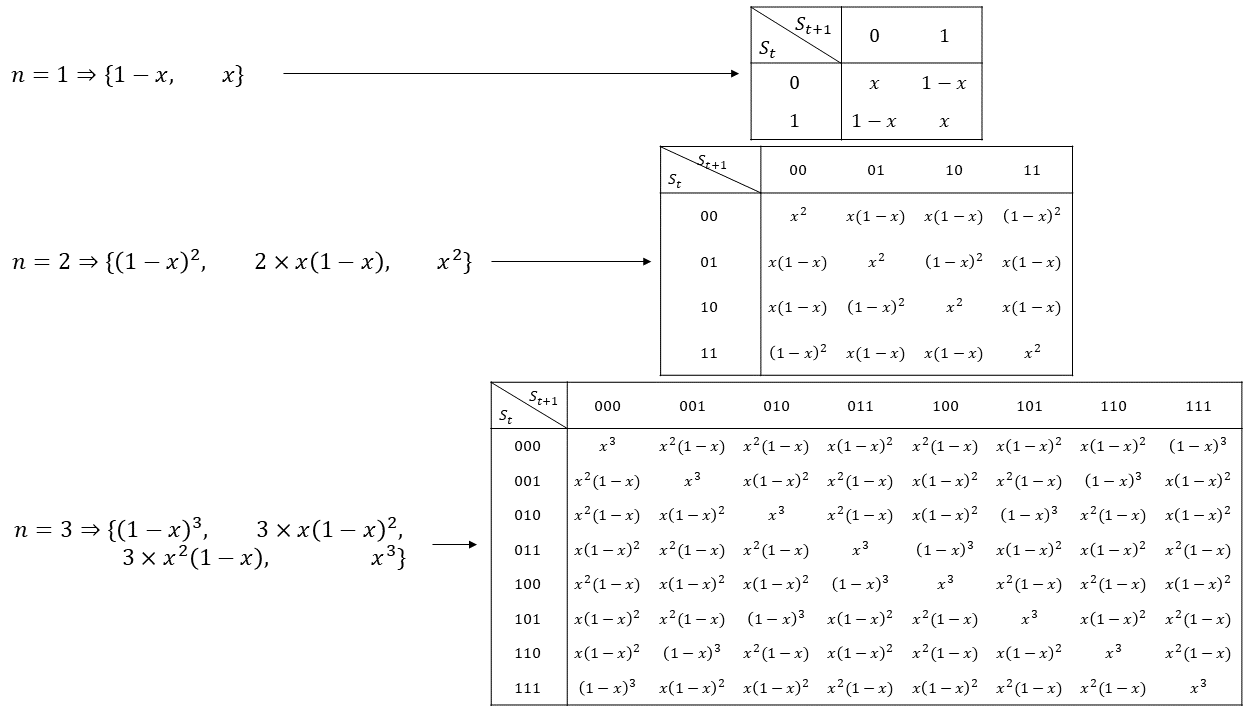}  
\caption{Row's distributions and synthetic TPMs of CMs with $n=1$, $n=2$, and $n=3$ scales , respectively.}
\label{reg_exp} 
\end{figure}

Firstly, we found row's regularities from synthetic TPMs, whose conditional probabilities are replaced with $x$ and $n$ polynomials by our framework. For instance, Figure \ref{reg_exp} illustrates row's distributions and synthetic TPMs of CMs with $n=1$, $n=2$, and $n=3$ scales, respectively. From one of three TPMs, each matrix's row's distribution is composed of concrete $x$ polynomials, such as every row of four-state TPM contains three kinds of items, $(1-x)^2$, $x(1-x)$, and $x^2$, and one scaled model's row distribution has different polynomial's kinds with other scaled models. According to polynomial's kinds of row's distributions of three scaled TPMs in Figure \ref{reg_exp}, we obtain Equation \ref{reg_polkinds}.

In the subsection \ref{equation_CQE}, we derived two generalization equations, Equations \ref{reg_polkinds} and \ref{reg_polnums}, by utilizing relationships among the layers of Pascal's Triangle \cite{rosen2007discrete} illustrated in Figure \ref{regularity}. These equations confirm two patterns in synthetic TPM's $row_i$: each row possesses the same quantity and types of $x$ polynomials, and the types and quantity of polynomials in each row depend on the current CM's variable amount $n$. To visually illustrate these regularities, Figure \ref{reg_exp} presents synthetic TPMs corresponding to CMs with $n=1$, $n=2$, and $n=3$ numbers of variables.

From the synthetic TPM's first regularity, featuring the same types and quantities of $x$ polynomials in $row_i$, we can initially simplify the computation of $D_{KL}(row_i||H^{max})$ \cite{edwards2008elements} to Equation \ref{dkl_new}. Furthermore, the relationship between polynomial types and quantities and the CM's variable number $n$ leads us to believe that Equation \ref{dkl_new} can be further simplified into an expression of $n$ and $x$. By utilizing $x$ polynomials of synthetic TPMs in Figure \ref{reg_exp} as conditional probabilities $p_i$ in $row_i$, we derived  three CM's $D_{KL}(row_i||H^{max})$ expressions, as shown in Equations \ref{new_kl_1}, \ref{new_kl_2}, and \ref{new_kl_3}.

\begin{align}
    \label{new_kl_1}
    D_{KL}(row_i||H^{max}) =&(1-x) * \log_2\left(\frac{(1-x)}{1/N}\right) + x * \log_2\left(\frac{x}{1/N}\right)\nonumber \\
                          =&(1-x) * \log_2\left(\frac{(1-x)}{1/2}\right) +  x * \log_2\left(\frac{x}{1/2}\right)\nonumber \\
                          =&(1-x) * \log_2\left(2*(1-x)\right) +  x * \log_2\left(2*x\right)\nonumber \\
                          =&\log_2(2)*[(1-x)+x] + (1-x) *\log_2\left(1-x\right) + x * \log_2\left(x\right)\nonumber \\
                          =&1 + (1-x) *\log_2\left(1-x\right) + x * \log_2\left(x\right)\nonumber \\
                          =&1 +(1-x)\log_2\left(1-x\right)+x\log_2\left(x\right)
\end{align}

\begin{align}
    \label{new_kl_2}
    D_{KL}(row_i||H^{max}) =&(1-x)^2 * \log_2\left(\frac{(1-x)^2}{1/N}\right) + 2* x(1-x) * \log_2\left(\frac{x(1-x)}{1/N}\right)\nonumber \\ 
                           &+ x^2 * \log_2\left(\frac{x^2}{1/N}\right)\nonumber \\
                          =&(1-x)^2 * \log_2\left(\frac{(1-x)^2}{1/4}\right) + 2* x(1-x) * \log_2\left(\frac{x(1-x)}{1/4}\right)\nonumber \\ 
                           &+ x^2 * \log_2\left(\frac{x^2}{1/4}\right)\nonumber \\
                          =&(1-x)^2 * \log_2\left(4*(1-x)^2\right) + 2* x(1-x) * \log_2\left(4*x(1-x)\right)\nonumber \\ 
                           &+ x^2 * \log_2\left(4*x^2\right)\nonumber \\
                          =&\log_2(4)*[(1-x)^2+2*x(1-x)+x^2] \nonumber\\&+ (1-x)^2 *\log_2\left((1-x)^2\right) +2* x(1-x) *\log_2\left(x(1-x)\right) + x^2 * \log_2\left(x^2\right)\nonumber \\
                          =&2 + 2*(1-x)^2 *\log_2\left(1-x\right) +2* x(1-x) *\log_2\left(1-x\right)\nonumber\\ &+2* x(1-x) *\log_2\left(x\right)+ 2*x^2 * \log_2\left(x\right)\nonumber \\
                          =&2 + 2*[(1-x)^2  + x(1-x) ]*\log_2\left(1-x\right)+2* [x(1-x) +x^2]* \log_2\left(x\right)\nonumber \\
                          =&2 + 2*(1-x)*\log_2\left(1-x\right)+2*x* \log_2\left(x\right)\nonumber \\
                          =&2*[1 +(1-x)\log_2\left(1-x\right)+x\log_2\left(x\right)]
\end{align}

\begin{align}
    \label{new_kl_3}
    D_{KL}(row_i||H^{max}) =&(1-x)^3 * \log_2\left(\frac{(1-x)^3}{1/N}\right) + 3* x(1-x)^2 * \log_2\left(\frac{x(1-x)^2}{1/N}\right)\nonumber \\ 
                           &+ 3* x^2(1-x) * \log_2\left(\frac{x^2(1-x)}{1/N}\right)+ x^3 * \log_2\left(\frac{x^3}{1/N}\right)\nonumber \\
                          =&(1-x)^3 * \log_2\left(\frac{(1-x)^3}{1/8}\right) + 3* x(1-x)^2 * \log_2\left(\frac{x(1-x)^2}{1/8}\right)\nonumber \\ 
                           &+ 3* x^2(1-x) * \log_2\left(\frac{x^2(1-x)}{1/8}\right)+ x^3 * \log_2\left(\frac{x^3}{1/8}\right)\nonumber \\
                          =&(1-x)^3 * \log_2\left(8*(1-x)^3\right) + 3* x(1-x)^2 * \log_2\left(8*x(1-x)^2\right)\nonumber \\ 
                           &+ 3* x^2(1-x) * \log_2\left(8*x^2(1-x)\right)+ x^3 * \log_2\left(8*x^3\right)\nonumber \\
                          =&\log_2(8)*[(1-x)^3+3*x(1-x)^2+3*x^2(1-x)+x^3] \nonumber\\&+ (1-x)^3 * \log_2\left((1-x)^3\right) + 3* x(1-x)^2 * \log_2\left(x(1-x)^2\right)\nonumber \\ 
                           &+ 3* x^2(1-x) * \log_2\left(x^2(1-x)\right)+ x^3 * \log_2\left(x^3\right)\nonumber \\
                          =&3  + 3*(1-x)^3 *\log_2\left(1-x\right) +6* x(1-x)^2 *\log_2\left(1-x\right)\nonumber\\ &+3* x^2(1-x) *\log_2\left(1-x\right)\nonumber\\ &+3* x(1-x)^2 *\log_2\left(x\right)+ 6*x^2(1-x) * \log_2\left(x\right)\nonumber \\&+3* x^3 *\log_2\left(x\right)\nonumber \\
                          =&3 + 3*[(1-x)^3  + 2*x(1-x)^2+x^2(1-x) ]*\log_2\left(1-x\right)\nonumber\\ &+3* [x(1-x)^2 + 2*x^2(1-x) +x^3]* \log_2\left(x\right)\nonumber \\
                          =&3 + 3*(1-x)*\log_2\left(1-x\right)+3*x* \log_2\left(x\right)\nonumber \\
                          =&3*[1 +(1-x)\log_2\left((1-x)\right)+x\log_2\left(x\right)]
\end{align}

Therefore, combining the simplified expressions for $D_{KL}(row_i||H^{max})$ provided by Equations \ref{new_kl_1}, \ref{new_kl_2}, and \ref{new_kl_3} with the two regularities of $row_i$, we summarize Equation \ref{reg_kl}, offering a direct calculation of CM's $D_{KL}(row_i||H^{max})$ from $n$ and $x$. Substituting Equation \ref{reg_kl} into Equation \ref{determinism} \cite{hoel2017map}, Equation \ref{reg_det} explicitly reveals the mathematical relationship between CM's $determinism$ and the framework parameter $x$. Furthermore, employing Equations \ref{EI_decop} and \ref{effectiveness} \cite{hoel2017map}, we derive the CQE equation, as demonstrated by Equation \ref{reg_cqe}, to optimize the computations of the quantification framework for calculating the $determinism$.

\section{Algorithms for quantifying the critical CE conditions}
\label{extensions}

In this section, we provide pseudo-codes as Algorithms \ref{TPM_solver}, \ref{solver}, and \ref{deg_gen} to describe the quantification processes of critical CE conditions within our quantification framework. Since we haven't simplified the computation of CM's $degeneracy$, these three algorithms require the synthetic TPMs to calculate the $degeneracy_{test}$ of the tested CM through Equation \ref{degeneracy} \cite{hoel2017map}. In the pseudo-codes, TPM\_Generator represents the generation process of synthetic TPMs corresponding to $deg\_vector$ in the test sample $[x,\ deg\_vector]$, as implemented by Algorithms \ref{implemt_alg_deg}, \ref{trans_st2va}, and \ref{trans_va2st} in Appendix \ref{algorithms_frame}. Figures \ref{quan_process1} and \ref{quan_process2} illustrate the execution processes of Algorithms \ref{TPM_solver} and \ref{solver}, respectively. Additionally, Algorithm \ref{deg_gen} is an auxiliary function block for the ANN. Therefore, unlike the other two quantification algorithms, it accepts the predicted $x$ from the ANN as input and directly uses $degeneracy_{target}$ as the conditional constraint to identify a specific $deg\_vector$ as the asymmetry condition for CE to occur.

Furthermore, if using the $EI_{real}$ and $n_{real}$ derived from Hoel's TPMs \cite{hoel2017map} as the inputs of Algorithms \ref{TPM_solver} and \ref{solver}, we can obtain the settings $[x,\ deg\_vector]$ of synthetic TPMs. This extensive application of the quantification algorithm can generate data as another part of the test set for validating the performance of trained ANNs.

\begin{algorithm}
\caption{TPM Solver}\label{TPM_solver}
\begin{algorithmic}
\State // \textit{input and initialize}
\State \textbf{Input:} $n_{test}$ and $EI_{target}$ \text{ \# Tested CM's variables and targeted EI}
\State \textbf{Output:} $x$ and $deg\_vector$ \text{ \# As critical CE conditions}
\State // \textit{implement}
\State tolerance = 1e-06 \text{ \# Set the error limitation of algorithm to $10^{-6}$}
\State deg\_vector\_s = [] \text{ \# All the possible $deg\_vector$s of $n$ scaled model}
\State \textbf{for} $FD$ in range(1, $2^{n-1}$): \text{ \# All the possible $FD$s of $n$ scaled model}
\State\quad \textbf{for} $\sum CD$ in range(1, $2^n$): \text{ \# All the possible $\sum CD$s of $n$ scaled mode}
\State\quad\quad \textbf{if} $\sum CD < 2FD$ and $FD>1$: \text{ \# Escape unreasonable values of $\sum CD$}
\State\quad\quad\quad continue
\State\quad\quad \textbf{else}:
\State\quad\quad\quad  deg\_vector\_s.append([$FD$, $\sum CD$]) \text{ \# Save the $deg\_vector$}
\State x\_s = linspace(1, 0.5, 1001) \text{ \# All the possible $x$s}
\State \textbf{for} deg\_vector in deg\_vector\_s: \text{ \# Search for satisfying $deg\_vector$}
\State\quad \textbf{for} x in x\_s: \text{ \# Search for satisfying $x$}
\State\quad\quad TPM\_s = TPM\_Generator(n, x, deg\_vector) \text{ \# Generate the TPM sets}
\State\quad\quad \textbf{for} TPM in TPM\_s:
\State\quad\quad\quad \text{ \# Calculate $determinism$ of synthetic TPM}
\State\quad\quad\quad $determinism_{test} =\frac{1}{N}\sum\limits_{row_i\in TPM}\frac{D_{KL}(row_i|H^{max})}{\log_2(N)}$
\State\quad\quad\quad \text{ \# Calculate $degeneracy$ of synthetic TPM}
\State\quad\quad\quad $degeneracy_{test}=\frac{D_{KL}\left(\sum\limits_{row_i\in TPM}row_i/N|I_D\right)}{\log_2(N)}$
\State\quad\quad\quad \text{ \# Calculate EI of synthetic TPM}
\State\quad\quad\quad $EI_{test} = n*(determinism_{test}-degeneracy_{test})$
\State\quad\quad\quad \textbf{if} $|EI_{target} - EI_{test}| < tolerance$: \text{ \# When error < $10^{-6}$}
\State\quad\quad\quad\quad \text{ \# Provide $x$ and $deg\_vector$ as conditions of uncertainty and asymmetry}
\State\quad\quad\quad\quad \textbf{return} $x$ and $deg\_vector$  
\end{algorithmic}
\end{algorithm}

\begin{algorithm}
\caption{CQE Solver}\label{solver}
\begin{algorithmic}
\State // \textit{input and initialize}
\State \textbf{Input:} $n_{test}$ and $EI_{target}$ \text{ \# Tested CM's variables and targeted EI}
\State \textbf{Output:} $x$ and $deg\_vector$ \text{ \# Settings of synthetic TPM}
\State // \textit{implement}
\State tolerance = 1e-06 \text{ \# Set the error limitation of algorithm to $10^{-6}$}
\State deg\_vector\_s = [] \text{ \# All the possible $deg\_vector$s of $n$ scaled model}
\State \textbf{for} $FD$ in range(1, $2^{n-1}$): \text{ \# All the possible $FD$s of $n$ scaled model}
\State\quad \textbf{for} $\sum CD$ in range(1, $2^n$): \text{ \# All the possible $\sum CD$s of $n$ scaled mode}
\State\quad\quad \textbf{if} $\sum CD < 2FD$ and $FD>1$: \text{ \# Escape unreasonable values of $\sum CD$}
\State\quad\quad\quad continue
\State\quad\quad \textbf{else}:
\State\quad\quad\quad  deg\_vector\_s.append([$FD$, $\sum CD$]) \text{ \# Save the $deg\_vector$}
\State x\_s = linspace(1, 0.5, 1001) \text{ \# All the possible $x$s}
\State \textbf{for} deg\_vector in deg\_vector\_s: \text{ \# Search for satisfying $deg\_vector$}
\State\quad \textbf{for} x in x\_s: \text{ \# Search for satisfying $x$}
\State\quad\quad TPM\_s = TPM\_Generator(n, x, deg\_vector) \text{ \# Generate the TPM sets}
\State\quad\quad \textbf{for} TPM in TPM\_s: \text{ \# Calculate $degeneracy$ }
\State\quad\quad\quad \text{ \# Calculate $determinism$ of synthetic TPM}
\State\quad\quad\quad $determinism_{test} = 1 + (1-x)*\log_2(1-x) + x*\log_2(x) $
\State\quad\quad\quad \text{ \# Calculate $degeneracy$ of synthetic TPM}
\State\quad\quad\quad $degeneracy_{test}=\frac{D_{KL}\left(\sum\limits_{row_i\in TPM}row_i/N|I_D\right)}{\log_2(N)}$
\State\quad\quad\quad \text{ \# Calculate EI of synthetic TPM}
\State\quad\quad\quad $EI_{test} = n*(determinism_{test}-degeneracy_{test})$
\State\quad\quad\quad \textbf{if} $|EI_{target} - EI_{test}| < tolerance$: \text{ \# When error < $10^{-6}$}
\State\quad\quad\quad\quad \text{ \# Provide $x$ and $deg\_vector$ as conditions of uncertainty and asymmetry}
\State\quad\quad\quad\quad \textbf{return} $x$ and $deg\_vector$  
\end{algorithmic}
\end{algorithm}

\begin{algorithm}
\caption{Vector Generator}\label{deg_gen}
\begin{algorithmic}
\State // \textit{input and initialize}
\State \textbf{Input:} $n_{test}$, $degeneracy_{target}$, and $x$
\State \text{ \# Tested CM's variables, ANN's prediction, and targeted $degeneracy$}
\State \textbf{Output:} $deg\_vector$ \text{ \# Correct $deg\_vector$}
\State // \textit{implement}
\State tolerance = 1e-06 \text{ \# Set the error limitation of algorithm to $10^{-6}$}
\State deg\_vector\_s = [] \text{ \# All the possible $deg\_vector$s of $n$ scaled model}
\State \textbf{for} $FD$ in range(1, $2^{n-1}$): \text{ \# All the possible $FD$s of $n$ scaled model}
\State\quad \textbf{for} $\sum CD$ in range(1, $2^n$): \text{ \# All the possible $\sum CD$s of $n$ scaled mode}
\State\quad\quad \textbf{if} $\sum CD < 2FD$ and $FD>1$: \text{ \# Escape unreasonable values of $\sum CD$}
\State\quad\quad\quad continue
\State\quad\quad \textbf{else}:
\State\quad\quad\quad  deg\_vector\_s.append([$FD$, $\sum CD$]) \text{ \# Save the $deg\_vector$}
\State \textbf{for} deg\_vector in deg\_vector\_s: \text{ \# Search for satisfying $deg\_vector$}
\State\quad TPM\_s = TPM\_Generator(n, x, deg\_vector) \text{ \# Generate the TPM sets}
\State\quad \textbf{for} TPM in TPM\_s: \text{ \# Calculate $degeneracy$ }
\State\quad\quad \text{ \# Calculate $degeneracy$ of synthetic TPM}
\State\quad\quad $degeneracy_{test}=\frac{D_{KL}\left(\sum\limits_{row_i\in TPM}row_i/N|I_D\right)}{\log_2(N)}$
\State\quad\quad \textbf{if} $|degeneracy_{target} - degeneracy_{test}| < tolerance$: \text{ \# When error < $10^{-6}$}
\State\quad\quad\quad \textbf{return} $deg\_vector$ \text{ \# Provide $deg\_vector$ as the result}
\end{algorithmic}
\end{algorithm}

\section{Details of Regression Networks}
\label{regression}
\renewcommand\thefigure{\thesection.\arabic{figure}}
\renewcommand\thetable{\thesection.\arabic{table}}
\renewcommand{\theequation}{D\arabic{equation}}
\setcounter{figure}{0} 
\setcounter{table}{0} 
\setcounter{equation}{0}
\setlength{\extrarowheight}{5pt}

\begin{figure}[t] 
\centering 
\includegraphics[scale=0.35]{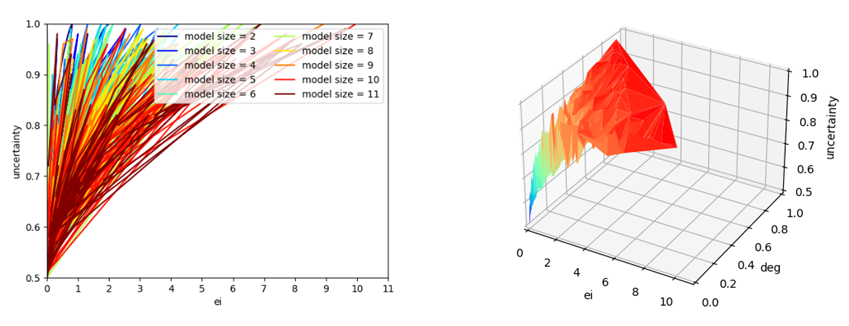}  
\caption{Illustrate the training dataset to show the relationships between $x$ and model's EI and between $x$, $degeneracy$ and EI, respectively.}
\label{data_ill} 
\end{figure}

To lower the computational complexity of our framework's quantification process achieved by Algorithm \ref{solver}, we used our framework to generate a dataset in Figure \ref{data_ill} for training a regression neural network to predict relative values of $x$ parameter of synthetic TPMs. As shown in Figure \ref{data_ill}, the relationship between the CM's $EI$ and the parameter $x$ is too complex to learn valuable features if the training data only uses the $EI$ and $n$ and the input features to the ANN. However, if considering the $degeneracy$ as another input feature of the neural network,  the relationship among the $EI$, $n$, $degeneracy$, and $x$ becomes more standard, as shown by the 3D surface in Figure \ref{data_ill}. Therefore, for the training data of the ANN, we chose the CM's variable numbers $n$, $degeneracy$, and $EI$ as input features to predict the value of the parameter $x$ as the numerical CE condition on the CM's uncertainty. Additionally, to validate the impact of data formats on the performance of the trained ANN,  we calculated negative, exponential, and logarithmical values of input features. Consequently, the regression network has six training sets containing various data formats, as denoted by the Orig., the Exp., the Log., the Neg\_Orig., the Neg\_Exp., and the Neg\_Log. to indicate the prediction results derived from which one of them.

\begin{figure}[b] 
\centering 
\includegraphics[scale=0.45]{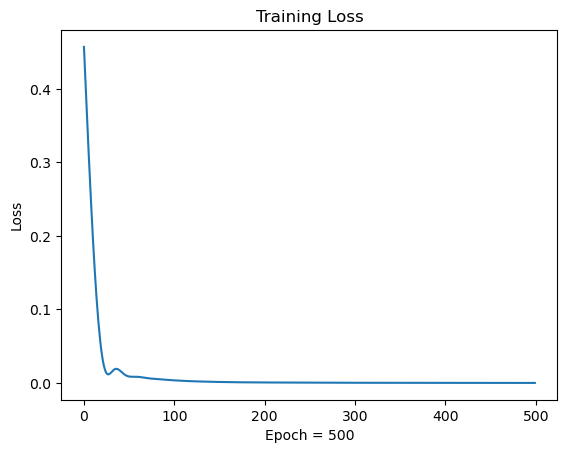}  
\caption{Illustrate the training dataset to show the relationships between $x$ and model's EI and between $x$, $degeneracy$ and EI, respectively.}
\label{training} 
\end{figure}

For the network's architecture, we chose a simple Fully Connected Network (FCN) \cite{goodfellow2016deep} as the fundamental structure of our regression network, as the FCN is not too large and deep to slow its prediction speed. In our experiments, we trained ten networks on data from CMs with different $n$ (our dataset contains ten values of $n$) to ensure the accuracy of $x$ predictions. Consequently, during the testing period, the corresponding network will be chosen to predict by the input feature $n$ in the test set. Finally, we trained every FCN on 360 training samples for 100, 500, and 1000 epochs and verified the network's performance on the test set with 40 samples. The performances of trained FCNs are measured by the Mean Squared Error (MSE) loss function, as demonstrated by Equation \ref{mse}. Figure \ref{training} displays the decrement of MSE loss to illustrate the process of training the networks by 500 epochs.

\begin{align}
    \label{mse}
    MSE = \frac{1}{n} \sum_{i=1}^{n} (y_i - \hat{y}_i)^2
\end{align}

Finally, we demonstrated FCNs' performances trained with data of six formats and ten different values of $n$ in Tables \ref{avg_perf_on_each_format} and \ref{Ivd_His_Perf_on_Neg_Orig_Format}. Meanwhile, the results in these tables are derived from testing trained networks on 40 samples generated by our framework to prove that the training process is correct. According to Table \ref{avg_perf_on_each_format},  the FCN trained by 500 epochs on the dataset with the Neg\_Orig. format provides the most accurate predictions, and Table \ref{Ivd_His_Perf_on_Neg_Orig_Format} supports that the networks trained by 500 epochs and 1000 epochs have similar performances on the test set. Therefore, we finally chose the FCN trained by 500 epochs on the Neg\_Orig. data format as the solution of solving the computational complexity by combining the network with Algorithm \ref{deg_gen} in Appendix \ref{extensions}.

\begin{table}[h]
\centering
\caption{Average Performances of FCNs Trained by Data in Six Formats}
\begin{tabular}{|c|c|c|c|c|c|c|}
\hline
\diagbox{Formats}{Models} & Orig. & Exp. & Log. & Neg\_Orig.   & Neg\_Exp. & Neg\_Log. \\ \hline
FCN\_100  & 0.0161   & 0.0150   & 0.5392    & 0.0128          & 0.0131        & 0.9363         \\ \hline
FCN\_500  & 0.0139   & 0.0103   & 0.0596    & \textbf{0.0029} & 0.0095        & 0.0646         \\ \hline
FCN\_1000 & 0.0138   & 0.0071   & 0.0326    & 0.0043          & 0.0132        & 0.0292         \\ \hline
\end{tabular}
\label{avg_perf_on_each_format}
\end{table}

\begin{table}[h]
\centering
\caption{Individual and Holistical Performances of FCNs Trained by Data in Neg\_Orig. Format}
\begin{tabular}{|c|cccccccccc|c|}
\hline
\multirow{2}{*}{Models} & \multicolumn{10}{c|}{Model Scales}                                                                                                                                                                                                                                                                                                                                        & \multirow{2}{*}{Avg.} \\ \cline{2-11}
                  & \multicolumn{1}{c|}{2}               & \multicolumn{1}{c|}{3}               & \multicolumn{1}{c|}{4}               & \multicolumn{1}{c|}{5}              & \multicolumn{1}{c|}{6}               & \multicolumn{1}{c|}{7}              & \multicolumn{1}{c|}{8}              & \multicolumn{1}{c|}{9}              & \multicolumn{1}{c|}{10}              & 11             &                       \\ \hline
FCN\_100          & \multicolumn{1}{c|}{0.06}          & \multicolumn{1}{c|}{0.02}          & \multicolumn{1}{c|}{0.02}          & \multicolumn{1}{c|}{3e-3}         & \multicolumn{1}{c|}{0.02}          & \multicolumn{1}{c|}{0.01}         & \multicolumn{1}{c|}{5e-3}         & \multicolumn{1}{c|}{3e-03}         & \multicolumn{1}{c|}{2e-3}          & 3e-3         & 0.01               \\ \hline
FCN\_500          & \multicolumn{1}{c|}{\textbf{0.01}} & \multicolumn{1}{c|}{\textbf{0.01}} & \multicolumn{1}{c|}{\textbf{0.01}}          & \multicolumn{1}{c|}{3e-4}         & \multicolumn{1}{c|}{\textbf{3e-4}} & \multicolumn{1}{c|}{3e-4}         & \multicolumn{1}{c|}{5e-3}         & \multicolumn{1}{c|}{1e-4}         & \multicolumn{1}{c|}{\textbf{3e-4}} & 1e-4        & \textbf{3e-3}       \\ \hline
FCN\_1000         & \multicolumn{1}{c|}{0.03}          & \multicolumn{1}{c|}{0.01}          & \multicolumn{1}{c|}{\textbf{0.01}} & \multicolumn{1}{c|}{\textbf{8e-5}} & \multicolumn{1}{c|}{\textbf{3e-4}} & \multicolumn{1}{c|}{\textbf{6e-5}} & \multicolumn{1}{c|}{\textbf{2e-5}} & \multicolumn{1}{c|}{\textbf{7e-5}} & \multicolumn{1}{c|}{\textbf{3e-4}} & \textbf{4e-5} & 4e-3                \\ \hline
\end{tabular}
\label{Ivd_His_Perf_on_Neg_Orig_Format}
\end{table}

\newpage
\section{GAP: An auxiliary functional block used by Algorithm \ref{implemt_alg_deg}}
\label{gap}
\renewcommand\thefigure{\thesection.\arabic{figure}}
\renewcommand\thetable{\thesection.\arabic{table}}
\renewcommand{\theequation}{D\arabic{equation}}
\setcounter{figure}{0} 
\setcounter{table}{0} 
\setcounter{equation}{0}
\setlength{\extrarowheight}{5pt}

\begin{algorithm}
\caption{GAP Block}\label{gap block}
\begin{algorithmic}
\State // \textit{inputs and output}
\State \textbf{Inputs:} $FD$ and $CD[0]-\text{ceil}\left(\frac{\sum CD}{FD}\right)$
\State \text{ \# The length of $CD$ array and the maximum variation for the first in $CD$ array.}
\State \textbf{Output:} $\Delta$
\State \text{ \# Possible variations of each number in $CD$ array.}
\State // \textit{implement}
\State \text{ \# Auxiliary function to generate all possible $CD$ arrays.}
\State \textbf{GENERATE\_HELPER}(current\_array, remaining\_length):
\State\quad \textbf{if} $\text{remaining\_length} = 0$:
\State\quad\quad  current\_array = current\_array[::-1]
\State\quad\quad  result.append(current\_array)
\State\quad \textbf{else}:
\State\quad\quad \textbf{if not} current\_array: 
\State\quad\quad\quad start\_value = 0  
\State\quad\quad \textbf{else}:
\State\quad\quad\quad start\_value = current\_array[-1]
\State\quad\quad \textbf{for} $i\in [start\_value, Delta + 1]$:
\State\quad\quad\quad \textbf{if} $i \leq CD[0]-\text{ceil}\left(\frac{\sum CD}{FD}\right)$:
\State\quad\quad\quad\quad \textbf{GENERATE\_HELPER}(current\_array + [i], remaining\_length - 1)
\State result = []
\State \textbf{GENERATE\_HELPER}([], $FD$)
\State $\Delta$ = result
\State \textbf{return} $\Delta$ 
\end{algorithmic}
\end{algorithm}

\bibliographystyle{ieeetr}  
\bibliography{paper}

\end{document}